\documentclass{article}

\usepackage{url}            
\usepackage{booktabs}       
\usepackage{amsfonts}       
\usepackage{nicefrac}       
\usepackage{microtype}      
\usepackage{xcolor}         
\usepackage{xspace}

\usepackage{subcaption}     
\usepackage{graphicx}  
\usepackage[colorlinks=true, linkcolor=blue, citecolor=blue, urlcolor=blue]{hyperref}
\usepackage{amsmath}
\usepackage{enumitem}
\usepackage{amsfonts}       
\usepackage{algpseudocode}
\usepackage{multirow}
\usepackage{float}

\newcommand{\system}{{MorphServe}\xspace}
\newcommand{\quant}{\small\textsf{LayerSwapper}\xspace}
\newcommand{\kvc}{\small\textsf{KVResizer}\xspace}

\newcommand{\FlashAttention}{\textsc{FlashAttention}\xspace} 
\newcommand{\Sarathi}{\textsc{Sarathi}\xspace} 
\newcommand{\PagedAttention}{\text{PagedAttention}\xspace}

\usepackage[accepted]{mlsys2025}

\mlsystitlerunning{MorphServe: Efficient and Workload-Aware LLM Serving via Runtime Quantized Layer Swapping and KV Cache Resizing}

\begin{document}

\twocolumn[
\mlsystitle{MorphServe: Efficient and Workload-Aware LLM Serving via Runtime Quantized Layer Swapping and KV Cache Resizing}




\begin{mlsysauthorlist}
\mlsysauthor{Zhaoyuan Su}{uva}
\mlsysauthor{Zeyu Zhang}{uva}
\mlsysauthor{Tingfeng Lan}{uva}
\mlsysauthor{Zirui Wang}{uva}
\mlsysauthor{Haiying Shen}{uva}
\mlsysauthor{Juncheng Yang}{harvard}
\mlsysauthor{Yue Cheng}{uva}
\end{mlsysauthorlist}

\mlsysaffiliation{uva}{University of Virginia}
\mlsysaffiliation{harvard}{Harvard University}


\mlsyskeywords{Machine Learning, MLSys}

\vskip 0.3in

\begin{abstract}
Efficiently serving large language models (LLMs) under dynamic and bursty workloads remains a key challenge for real-world deployment. Existing serving frameworks and static model compression techniques fail to adapt to workload fluctuations, leading to either service-level objective (SLO) violations under full-precision serving or persistent accuracy degradation with static quantization. To deal with these issues, we present {\system}, a dynamic, workload-aware LLM serving framework based on \emph{morphological adaptation}. {\system} introduces two asynchronous, token-level runtime mechanisms: \emph{quantized layer swapping}, which selectively replaces less impactful layers with quantized alternatives during high-load periods, and \emph{pressure-aware KV cache resizing}, which dynamically adjusts KV cache capacity in response to memory pressure. These mechanisms enable state-preserving transitions with minimum runtime overhead and are fully compatible with modern scheduling and attention techniques. Extensive experiments on Vicuna and Llama family models with real-world workloads demonstrate that {\system} reduces average SLO violations by 92.45\% and improves the P95 TTFT latency by $2.2$-\emph{$3.9\times$} compared to full-precision serving, without compromising generation quality. These results establish {\system} as a practical and elastic solution for LLM deployment in dynamic environments. 
\end{abstract}
]

\printAffiliationsAndNotice{}  

\section{Introduction}
\label{sec:intro}

The rise of large language models (LLMs) has made efficient and reliable serving a core challenge in modern AI infrastructure. Systems like vLLM~\cite{kwon2023efficient_vllm} and Orca~\cite{yu2022orca} optimize throughput via {\PagedAttention}~\cite{kwon2023efficient_vllm} and continuous batching~\cite{yu2022orca, sun2024llumnix, he2024deferred_llm_batching}, but assume fixed-precision execution and stable workloads. In contrast, real-world LLM workloads are dynamic and bursty~\cite{wang2024burstgpt, patel2024splitwise, azure_LLM_inference_trace}, with fluctuating request rates and context lengths. Even brief load spikes can cause memory exhaustion or queueing delays, leading to SLO violations---such as increased time-to-first-token (TTFT) and time-per-output-token (TOPT)---that degrade both user experience and system throughput. 

One naive solution is to statically over-provision GPU resources to accommodate worst-case traffic spikes. 
However, over-provisioning leads to substantial cost inefficiencies during underutilized periods~\cite{jaiswal2025serving_provisioning, fu2024serverlessllm}. Moreover, edge deployments lack the flexibility for dynamic scaling altogether~\cite{cai2024edge_llm}. Thus, the inability to elastically match model resource usage to real-time demand results in either SLO violations under pressure, or significant resource waste during low-load intervals.

Model compression techniques, such as quantization~\cite{lin2024awq, frantar2022gptq, lin2024duquant, shao2023omniquant}, pruning~\cite{ma2023llm_pruning, sreenivas2024llm_pruning, gao2024disp_pruning}, or low-rank approximation~\cite{hu2022lora, wu2024dlora}, offer an alternative approach by statically reducing the resource footprint of deployed LLMs. While these methods are effective in lowering memory and compute demands, they introduce irreversible accuracy degradation that persists even during periods of low load, when full-precision inference could be served without penalty. This results in a rigid, suboptimal quality--efficiency tradeoff that fails to align with workload variability. Key-value cache (KVC) compression~\cite{cai2024pyramidkv, zhou2024dynamickv, li2024snapkv} and eviction~\cite{liu2023scissorhands, feng2024ada} methods have been proposed to further reduce memory usage. However, these techniques often rely on fixed heuristics, cannot adapt to different workloads, lack compatibility with modern attention variants like Grouped Query Attention (GQA)~\cite{ainslie2023gqa, chen2024optimised_gqa, chinnakonduru2024weighted_gqa} and Multi-Head Latent Attention (MLA)~\cite{meng2025transmla, liu2024deepseek}, and remain inflexible to runtime serving conditions.

To handle dynamic workloads and address the above issues, in this paper, we present {\system}, a dynamic, workload-aware LLM serving framework based on morphological adaptation. {\system} continuously monitors system load and morphs model components—transformer layers and KVC blocks—on the fly in response to real-time memory pressure. When resource usage surges, {\system} reduces model footprint by replacing selected full-precision layers with lightweight quantized alternatives and expands KVC capacity by dynamically attaching additional memory blocks. These adaptations are reversed as pressure subsides, restoring full precision and reclaiming memory from KVC without interrupting inference.

{\system} contributes the following:
(1) A \textit{runtime layer swapping} mechanism that enables \textit{workload-aware mixed-precision serving}, allowing quantized and full-precision layers to coexist and be dynamically reconfigured based on runtime pressure without model flushing or architectural changes.
(2) A \textit{pressure-aware KVC resizing} mechanism that elastically adjusts KV cache capacity, supporting efficient batch prefilling and decoding under bursty traffic.
(3) A tunable runtime policy that \textit{navigates the accuracy–latency Pareto frontier}, balancing high-fidelity and low-latency objectives.
(4) Full compatibility with existing KVC compression and eviction schemes, enabling further efficiency gains with minimal accuracy degradation.

To achieve this, {\system} introduces two complementary morphing mechanisms, both designed to support asynchronous and compatible kernel executions with minimal overhead:
{\quant} identifies low-impact transformer layers by a sensitivity-based profiling, selectively and asynchronously replacing them with lower-precision alternatives at runtime. 
{\kvc} adaptively adjusts KVC capacity under memory pressure and runs in parallel with decoding using separate CUDA streams, ensuring seamless execution. 


Across extensive experiments on Llama 2, Llama 3, CodeLlama, and Vicuna using four datasets~\cite{gov_report, zhong2021qmsum, he2017dureader, fabbri2019multi} under Azure LLM Inference~\cite{azure_LLM_inference_trace} and BurstGPT~\cite{wang2024burstgpt} traces, {\system} reduces average SLO violations by 92.45\% and P95 TTFT latency by 2.2$\times$–3.9$\times$ over full-precision serving, while preserving comparable accuracy. Compared to static quantization via AWQ~\cite{lin2024awq}, {\system} reduces F1 and Rouge-L degradation by up to 88.85\% and improves memory utilization by 29.29\%. These results demonstrate {\system}'s ability to adapt to dynamic workloads while balancing latency and accuracy.

\section{Related Work}

\noindent\textbf{LLM Serving Systems.}
TorchServe~\cite{torchserve} and NVIDIA Triton~\cite{triton} provide general-purpose serving frameworks for parallel inference. 
Recent systems specialize in LLM serving~\cite{fast_transformer, tgi}, improving scheduling and memory management. 
Orca~\cite{yu2022orca} introduces continuous batching, and vLLM~\cite{kwon2023efficient_vllm} proposes PagedAttention, both aimed at improving throughput. 
{\Sarathi}~\cite{agrawal2023sarathi} and FastServe~\cite{wu2023fastserve} further reduce queuing delays via chunked prefill or preemptive scheduling. 
Despite these advances, existing systems assume fixed precision and static memory provisioning, which limits their responsiveness to bursty workloads. 
In contrast, {\system} enables fully online, workload-aware adaptation through runtime quantized layer swapping and elastic KV cache resizing, achieving smooth performance-accuracy tradeoffs under overload.

\noindent\textbf{LLM Post-Training Quantization.}
Post-training quantization (PTQ) reduces inference cost by compressing model weights~\cite{dettmers2023spqr, kim2023squeezellm, li2024norm_quant, zhao2024atom, lin2024awq, frantar2022gptq, xiao2023smoothquant, lin2024duquant}. 
Round-to-nearest schemes~\cite{yao2023comprehensive_quant, nagel2021white_paper_quantization} are simple but inaccurate for outlier-heavy layers. 
Non-uniform and calibration-based methods, including AdaRound~\cite{nagel2020up_adaround}, ZeroQuant~\cite{yao2022zeroquant}, GPTQ~\cite{frantar2022gptq}, and AWQ~\cite{lin2024awq}, achieve higher fidelity via adaptive rounding or activation statistics. 
Activation quantization has also been explored (e.g., SmoothQuant~\cite{xiao2023smoothquant}, QServe~\cite{lin2024qserve}, DuQuant~\cite{lin2024duquant}), but remains static and incurs irreversible degradation even when resources permit full precision. 
{\system} provides a systematic solution that dynamically switches between different precision layers based on real-time service pressure, compatible with any weight quantization method, such as GPTQ and AWQ.

\noindent\textbf{{Mixed-Precision and Adaptive-Depth Inference.}}
Mixed-precision execution improves efficiency by leveraging lower-bit operations~\cite{reggiani2023mix_precision, huang2021mixed_precision, wang2020towards_bitsplit, chen2024progressive_quant, frantar2025marlin_mixed_precision, dong2019hawq_mixed_precision, risso2022channel_mixed_precision}. 
Hardware-level methods (Mix-GEMM~\cite{reggiani2023mix_precision}, Bit-Split~\cite{wang2020towards_bitsplit}) and prompt-aware schemes (PMPD~\cite{chen2024progressive_quant}, MARLIN~\cite{frantar2025marlin_mixed_precision}) adjust precision based on token entropy or pre-computed sensitivity. 
{Recent adaptive-depth and early-exit frameworks (e.g., FlexiDepth~\cite{luo2025adaptive}, LayerSkip~\cite{Elhoushi_2024}) reduce computation via layer skipping but require architectural modification and lack real-time adaptation.} 
{In contrast, {\system} achieves dynamic, workload-aware mixed-precision serving guided by runtime telemetry, coupling token-level precision with system load for smooth efficiency--accuracy balancing.}

\section{Motivation}
\label{sec:motivation}

\begin{figure*}[t]
    \centering
        \begin{subfigure}[t]{0.24\textwidth}
        \includegraphics[width=\linewidth]{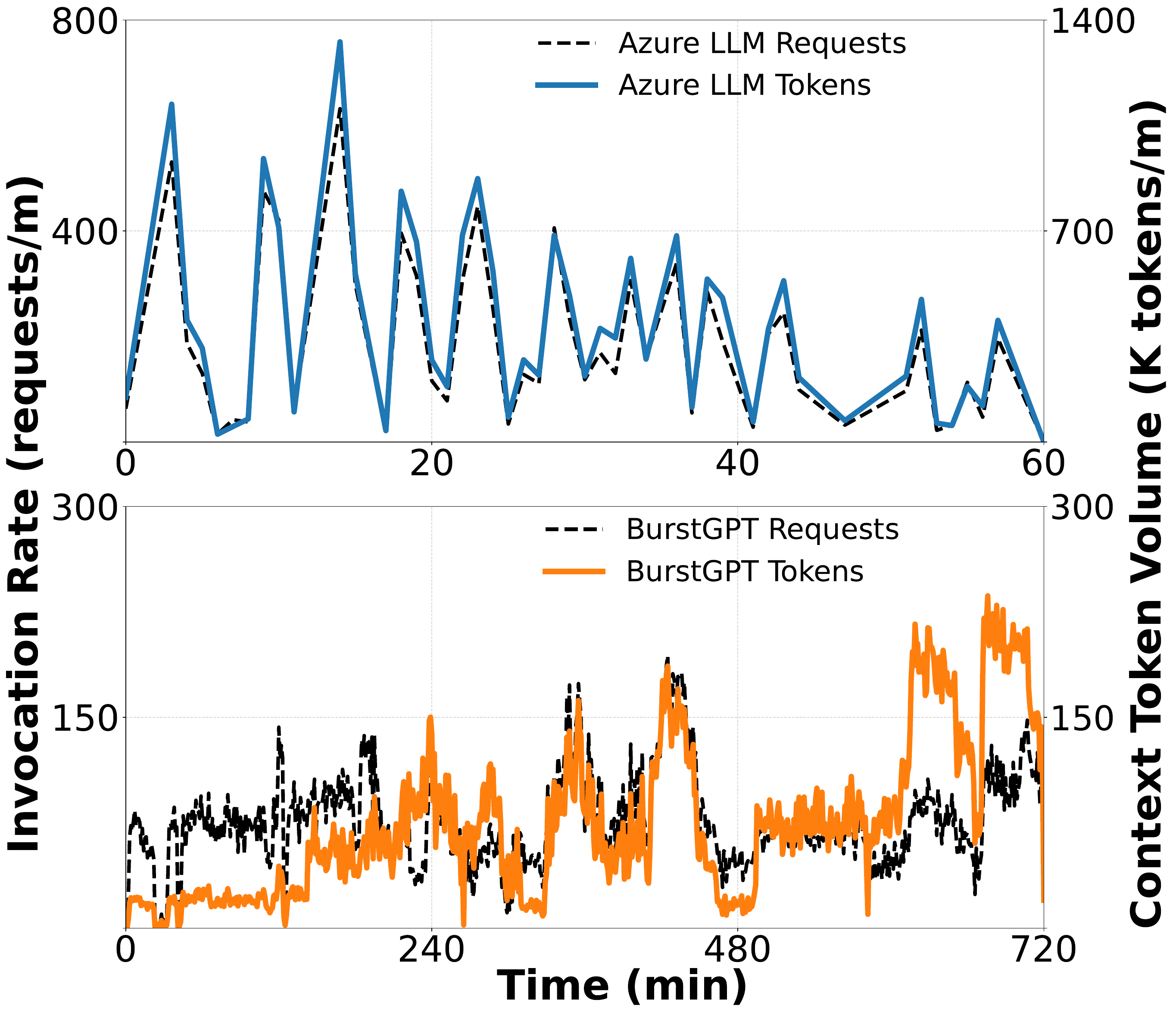}
        \caption{}
        \label{fig:workloads}
        \end{subfigure}
        \begin{subfigure}[t]{0.24\textwidth}
        \includegraphics[width=\linewidth]{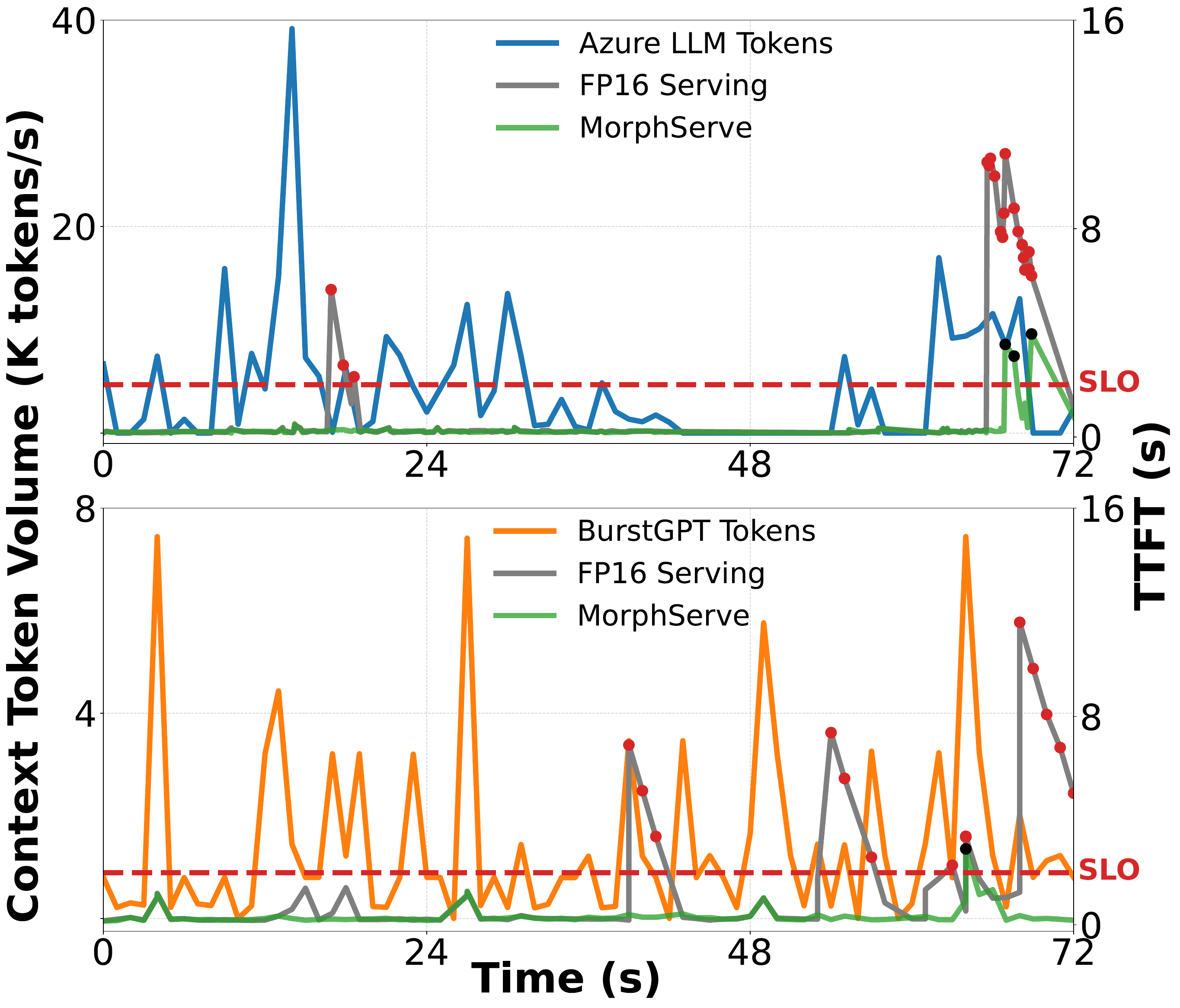}
        \caption{}
        \label{fig:saturation_point}
        \end{subfigure}
        \begin{subfigure}[t]{0.24\textwidth}
        \includegraphics[width=\linewidth]{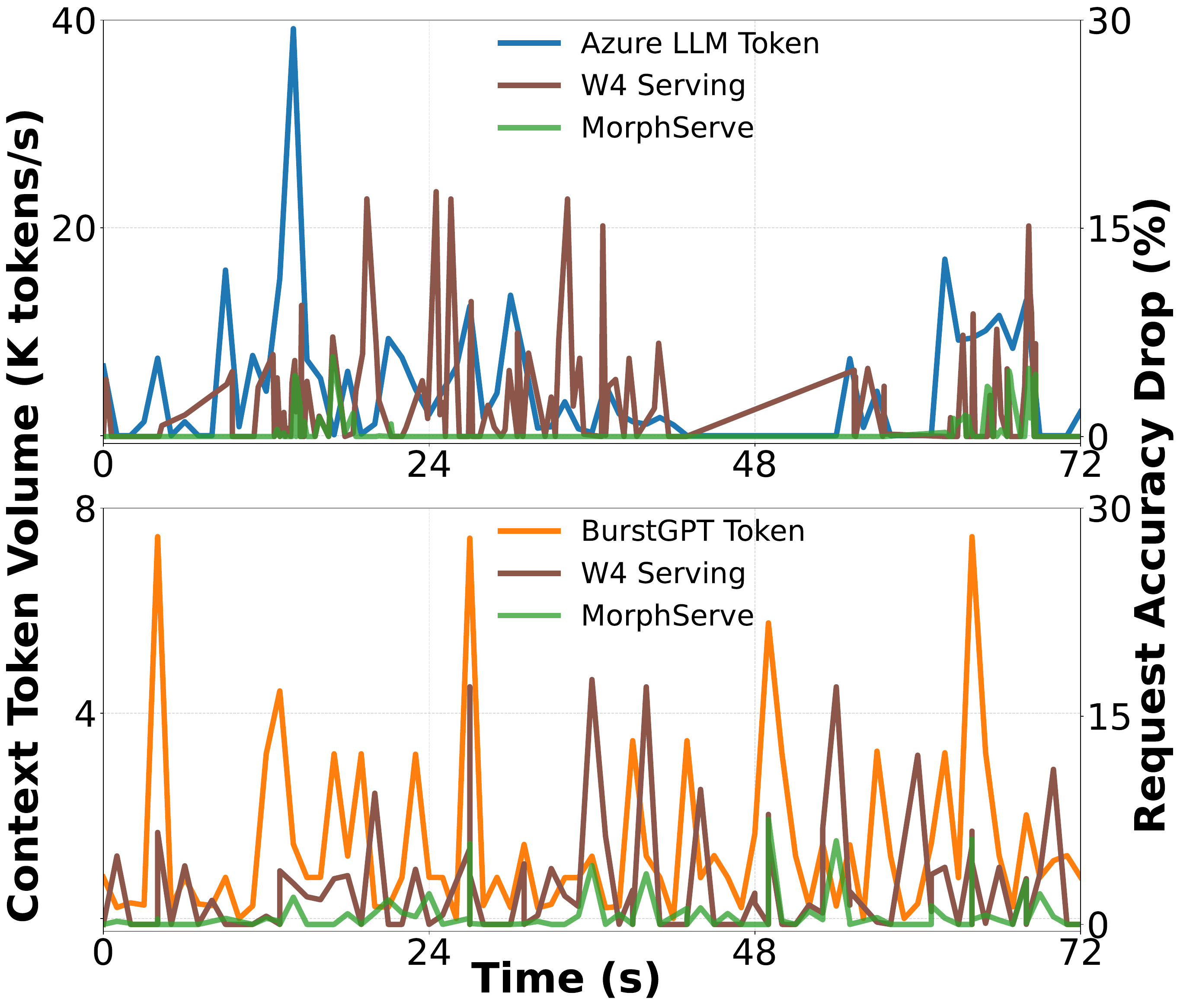}
        \caption{}
        \label{fig:static_compression}
        \end{subfigure}
        \begin{subfigure}[t]{0.23\textwidth}
        \includegraphics[width=\linewidth]{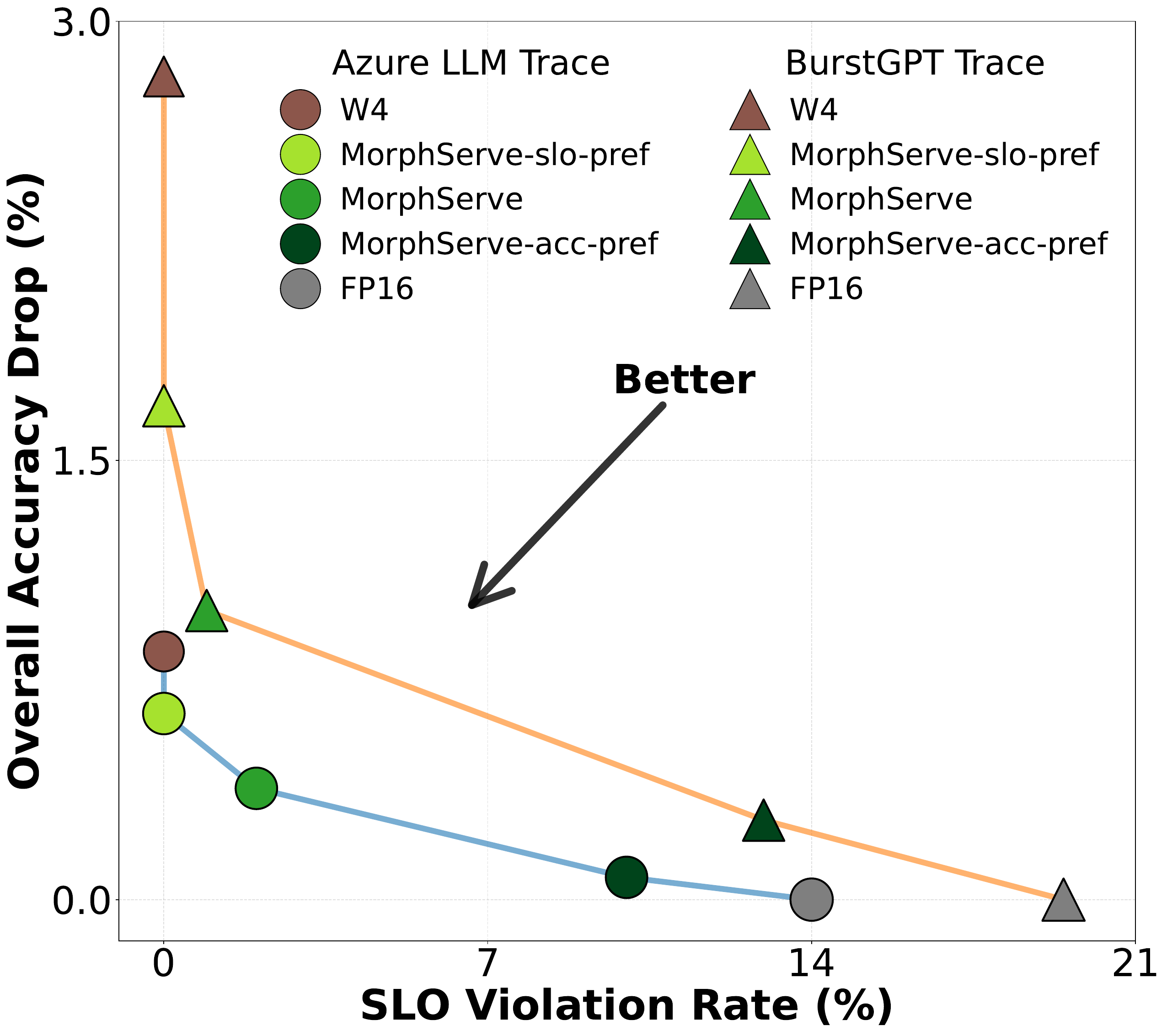}
        \caption{}
        \label{fig:parato_tradeoff}
        \end{subfigure}
    \vspace{-5pt}
    \caption{
    \textbf{Motivation for dynamic adaptation design in LLM serving.}
    (a) Real-world LLM workloads are highly dynamic and bursty in request and token volume.
    (b) Full-precision serving suffers TTFT spikes and SLO violations when workload exceeds the saturation point.
    (c) Statically quantized model causes constant accuracy degradation even during low-load periods when it is possible to serve full-precision models. 
    (d) {\system} dynamically adapts to resource pressure and consistently achieves an optimal balance between SLO compliance and accuracy.}
    \vspace{-15pt}
    \label{fig:motivations}
\end{figure*}

\textbf{Real-world LLM workloads are highly bursty.}
LLM serving systems face highly dynamic and bursty traffic patterns in real-world scenarios. As shown in \hyperref[fig:workloads]{Figure~\ref{fig:workloads}}, the production
workloads of Microsoft Azure LLM services~\cite{azure_LLM_inference_trace, stojkovic2025dynamollm} and BurstGPT~\cite{wang2024burstgpt} reveal rapid fluctuations in both the request arrival rates (i.e., request bursts) and the volumes of tokens. These fluctuations reflect the \textit{non-stationary nature} of practical LLM inference workloads, which deviates from the traditional assumptions of most serving schemes~\cite{kwon2023efficient_vllm, yu2022orca, jaiswal2025serving_provisioning}.

\textbf{Request burst leads to long TTFT and SLO violation.}
As system load increases, even small surges can cause sharp spikes in TTFT latency. In this work, we set the TTFT SLO threshold to 2~seconds, consistent with prior work~\cite{xiong2024layerkv, gao2024cost_cachedattention, qiao2024conserve}. As shown in \hyperref[fig:saturation_point]{Figure~\ref{fig:saturation_point}}, full-precision serving quickly exceeds the SLO threshold once it reaches the \textbf{saturation point}: defined as \textit{the load level at which GPU memory becomes insufficient to schedule new requests for prefilling or to continue decoding for the ongoing batch}. 
At this point, incoming requests are forced to wait until memory is reclaimed, \textit{incurring significant queueing latency with SLO violation}.  

\textbf{Static quantization trades quality for efficiency irrespective of load.} 
To mitigate resource constraints, static quantization methods~\cite{lin2024awq,frantar2022gptq,lin2024duquant} have been widely adopted. However, these methods introduce persistent accuracy degradation across all conditions, regardless of whether the system is overloaded.
As shown in \hyperref[fig:static_compression]{Figure~\ref{fig:static_compression}}, the INT4 quantized model with AWQ~\cite{lin2024awq} consistently degrades accuracy, measured by F1 score following~\cite{lin2024duquant}, on the {\small\texttt{GovReport}} dataset from LongBench~\cite{bai2023longbench}, even during periods when full-precision inference is feasible. This demonstrates that static quantization over-prioritizes efficiency, sacrificing quality during low-load intervals.

\textbf{Workload-aware adaptation achieves optimal tradeoffs.}
As shown in the Pareto analysis in \hyperref[fig:parato_tradeoff]{Figure~\ref{fig:parato_tradeoff}}, {\system} achieves superior tradeoffs by aligning dynamic quantization with real-time workload conditions.
{A key insight in MorphServe is \textit{adaptive mixed-precision LLM serving}, where quantized and full-precision layers can coexist and are dynamically reconfigured.}
This contrasts with static quantization and recent dynamic methods~\cite{chen2024progressive_quant, frantar2025marlin_mixed_precision, reggiani2023mix_precision}, which prioritize serving performance or hardware efficiency but overlook runtime workload variability.
Most importantly, {\system} enables smooth navigation along the accuracy---efficiency Pareto frontier, from uncompressed, high-accuracy models to highly quantized, efficient ones.

\section{System Design}
\label{sec:design}

{\system} is designed with three primary objectives: 
(1)~Dynamic adaptation: Respond to real-time workload demands and GPU memory pressure by dynamically adjusting model layer configuration and KVC capacity on the fly during inference.
(2)~Accuracy preservation: Ensure no degradation under light or moderate load, and introduce only minimal, necessary, and fine-grained token-level accuracy loss to sustain serving performance beyond the saturation point. 
(3)~Low overhead: Minimize the performance impact of dynamic adaptation by leveraging asynchronous execution and overlapping. 

\subsection{Architecture and Workflow}

\begin{figure*}[t]
    \centering
    \includegraphics[width=0.78\textwidth]{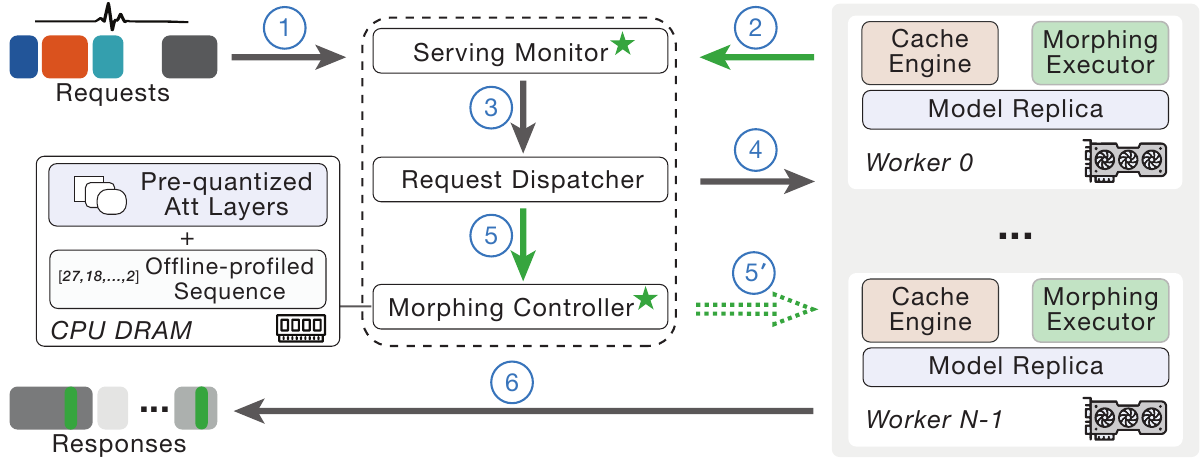}
    \caption{
    \textbf{{MorphServe} dynamic adaptation workflow.} 
    Incoming requests (1) and real-time telemetry from workers (2) are aggregated by the Serving Monitor and sent to the Request Dispatcher (3).
    The Dispatcher routes requests to workers (4) and forwards runtime metrics to the Morphing Controller (5), which detects resource pressure and issues adaptation commands (5').
    Responses (6) are returned to users, with only a small portion of tokens (in green) generated by mixed-precision layers.
    }
    \label{fig:system_design}
    \vspace*{-1.25em} 
\end{figure*}

\noindent\textbf{System Architecture.}  
As illustrated in \hyperref[fig:system_design]{Figure~\ref{fig:system_design}}, {\system} consists of three core components—\textit{Serving Monitor}, \textit{Morphing Controller}, and \textit{Morphing Executor}, which together form a feedback-driven control loop for dynamic adaptation. 

\begin{itemize}[noitemsep,leftmargin=*]
\vspace*{-0.5em}

    \item \textit{Serving Monitor} collects runtime metrics from all workers, including GPU memory utilization, request queue depth, throughput, and token-level latency (TTFT and TPOT). These metrics are smoothed over short time windows to identify workload shifts and early signs of system saturation. 
    
    \item \textit{Morphing Controller} serves as the global GPU memory manager. When monitored metrics exceed user-predefined thresholds (e.g., KVC memory usage over 85\%, queueing delay over 100 ms), 
    it decides whether to trigger \textit{selective layer swapping} (\hyperref[sec:runtime_swapping]{Section~\ref{sec:runtime_swapping}}) and \textit{elastic KVC resizing} (\hyperref[sec:kv_morphing]{Section~\ref{sec:kv_morphing}}),
    and dispatches corresponding instructions to the target workers.   
    
    \item \textit{Morphing Executor} resides on each worker and executes adaptation commands locally. It dynamically reconfigures the model using {\quant} (\hyperref[sec:runtime_swapping]{Section~\ref{sec:runtime_swapping}}), which switches a selective set of layers between full-precision and pre-quantized layers, or between different quantization levels (e.g., from INT8 to INT4) to reduce resource usage and improve inference latency under pressure. In addition, it applies {\kvc} (\hyperref[sec:kv_morphing]{Section~\ref{sec:kv_morphing}}) to adjust KVC memory allocation by elastically expanding or shrinking the number of KVC blocks as needed. 
    All adaptations are asynchronous overlapping communication and computation~\cite{async_cuda_stream} with preallocated memory buffers to seamlessly overlap with ongoing inference.  

\vspace*{-0.5em}
\end{itemize}

{\system}'s adaptive and versatile architecture enables efficient and timely operation across diverse and bursty workloads, ensuring serving quality under pressure while avoiding unnecessary degradation during underloaded periods. 

\noindent\textbf{Token-level Workload 
Adaptation.}
Unlike existing model and KVC compression schemes, which affect  
the entire request~\cite{model_compression_2018_iclr, lin2024awq, frantar2022gptq, lin2024duquant, zhao2024atom}, {\system} enables \emph{fine-grained, token-level} workload adaptation. 
During a single request’s decoding phase, {\system} may temporarily replace a subset of layers. For example, switching 2 layers from full-precision to INT4 when saturation is detected (examples shown in \hyperref[sec:runtime_swapping]{Section~\ref{sec:runtime_swapping}}). This allows early tokens to be generated at full precision, while only later tokens experience minimal accuracy degradation. Once the pressure subsides, the affected layers are restored to full precision, enabling continued decoding at the original accuracy. As a result, accuracy degradation is confined to a small portion of tokens, even within a single request. 

\noindent\textbf{State-Preserving Morphing During Inference.}
A key feature of {\system}’s serving workflow is its ability to seamlessly adapt model layer precision and elastically resize KVC capacity on-the-fly during request execution, 
\emph{without model flushing or re-prefilling}. 
When system pressure triggers adaptation, the Morphing Controller can selectively swap model layers without disrupting the attention state or decoding progress, avoiding expensive \textit{serving pauses} and \textit{recomputation}. 
This design allows {\system} to intervene mid-inference at the token level, preserving continuity in generation and enabling real-time adaptation with minimal runtime interference. 

\subsection{Offline Profiling for Layer Swapping Sequence}
\label{sec:offline_profiling}

To identify a layer swapping sequence that minimizes accuracy impact during runtime, {\system} performs \textit{offline profiling} to construct a prioritized swapping order based on sensitivity analysis. In this subsection, we describe how {\system} profiles and ranks layers to establish this sequence with a focus on accuracy and robustness.

\noindent\textbf{Problem Statement.} 
The objective is to minimize cumulative accuracy degradation over the time interval during which one or more layers are quantized.

Let $f(x_t)$ denote the full-precision model output at time $t$, and $f^{(Q_t)}(x_t)$ the output when a subset of layers $Q_t$ are quantized at that time. The cumulative degradation over the interval $[t_1, t_n]$ can be formulated as: \vspace{-20pt}

\begin{equation}
\min_{\{n_k\}} \sum_{t = t_1}^{t_n} \Delta(f(x_t), f^{(Q_t)}(x_t))
\label{eq:objective}
\end{equation}

This problem has a sequential and state-dependent structure: each swapping decision impacts downstream accuracy until the corresponding layer is restored to full precision. Selecting which layers to replace introduces combinatorial complexity, making exact optimization intractable. To address this, {\system} performs offline profiling using hybrid sensitivity metrics to evaluate the accuracy impact of each layer. The resulting sequence provides a prioritized order of layers that can be replaced with minimal expected accuracy degradation. We now describe the sensitivity metrics and the greedy policy used to construct this sequence.

\noindent\textbf{Independence of Layer Quantization Effects.}
To validate the independence of layer-wise quantization effects, we incrementally quantized the early layers of Llama 2 7B. We observed that the additional perplexity from quantizing a layer remained nearly constant across all settings. This supports the assumption that quantization impact is approximately additive across layers. For detailed results, please refer to \hyperref[appendix:layer_independence]{Appendix} Table 4.

\noindent\textbf{Sensitivity Analysis for Layer Swapping.}
To construct the swapping sequence, {\system} estimates the sensitivity of each decoder layer using cosine similarity-based local and global metrics that capture its impact on overall model accuracy. These sensitivity scores are used to rank layers, providing a prioritized order that approximates the optimal swapping strategy. \vspace{-5pt}

\begin{itemize}[leftmargin=*]
    \item \textit{Layer Transformation Sensitivity (LTS)} measures the direct change between a layer’s input and output: \vspace{-5pt}
    \begin{equation}
    \text{LTS}_p = \cos\left(h_p(x), x_p\right)
    \label{eq:IOA}
    \end{equation}
    where \( x_p \) is the input and \( h_p(x) \) is the output of layer \( p \). Lower similarity indicates stronger transformations and higher potential sensitivity to layer swapping.

    \item \textit{Layer Replacement Sensitivity (LRS)} quantifies the output distortion caused by replacing the original layer with its quantized version:\vspace{-5pt}
    \begin{equation}
    \text{LRS}_p = \cos\left(h_p(x), h_p^Q(x)\right)
    \label{eq:MES}
    \end{equation}
    where \( h_p^Q(x) \) is the output of layer \( p \) with quantized weights. Lower similarity implies greater deviation due to replacement. 

    \item \textit{Model Degradation Sensitivity (MDS)} measures the model-level accuracy impact from replacing a layer \( p \) given the current set of quantized layers \( Q \):\vspace{-5pt}
    \begin{equation}
    \text{MDS}_p^{(Q)} = \cos\left(f^{(Q)}(x), f^{(Q \cup \{p\})}(x)\right)
    \label{eq:TLI}
    \end{equation}
    where \( f^{(Q)}(x) \) is the model output with layers \( Q \) replaced. This state-aware metric captures the incremental global degradation introduced by swapping layer \( p \) in the current context. \vspace{-5pt}
\end{itemize}

We combine these metrics into a unified \textbf{Layer Importance Score (LIS)}: 

\begin{equation}
\text{LIS}_p = \alpha_1 \cdot \text{LTS}_p + \alpha_2 \cdot \text{LRS}_p + \beta \cdot \text{MDS}_p^{(Q)}
\label{eq:LIS}
\end{equation}

In this formulation, $\mathrm{LTS}_p$ and $\mathrm{LRS}_p$ are \textit{local sensitivity metrics} that evaluate the behavior of the layer $p$ in isolation, while $\mathrm{MDS}^{(Q)}_p$ is a \textit{global metric} that measures the model-level degradation when replacing $p$, given the current replaced layer set $\mathcal{Q}$. For a given model, the LIS for each layer is computed \textit{offline during profiling}, and the resulting sequence is stored and used directly at runtime. This design avoids any runtime recomputation or decision-making overhead.
We use a weighted combination of three sensitivity metrics to compute the LIS: weight sensitivity ({$\alpha_1$}), quantization distortion ({$\alpha_2$}), and end-to-end degradation ({$\beta$}), with $\alpha_1=\alpha_2=0.25$ and $\beta=0.5$. This configuration emphasizes stability and user-perceived quality while retaining fine-grained layer-local signals. Compared to a uniform weighting scheme, our setting yields more consistent layer rankings and improved performance. Full results and ablations appear in {Appendix} Table 1.

\noindent\textbf{Metric Choice in Sensitivity Computation.}
{We use cosine similarity instead of the L2 norm to assess layer sensitivity,
as it captures direction-based shifts across transformer layers and mitigates
scale noise. This choice follows analyses of layer relevance in
Transformers~\cite{whatmatters2025,transformer-layers-as-painters,jiang2025tracing}.
Cosine-based LIS yields lower perplexity than L2 variants across quantization levels, confirming its suitability for layer profiling. Our detailed evaluation comparison is in {Appendix} Table 2.

\subsection{{\quant}: Runtime Layer Swapping}
\label{sec:runtime_swapping}

\begin{figure*}[t]
    \centering
    \includegraphics[width=0.98\textwidth]{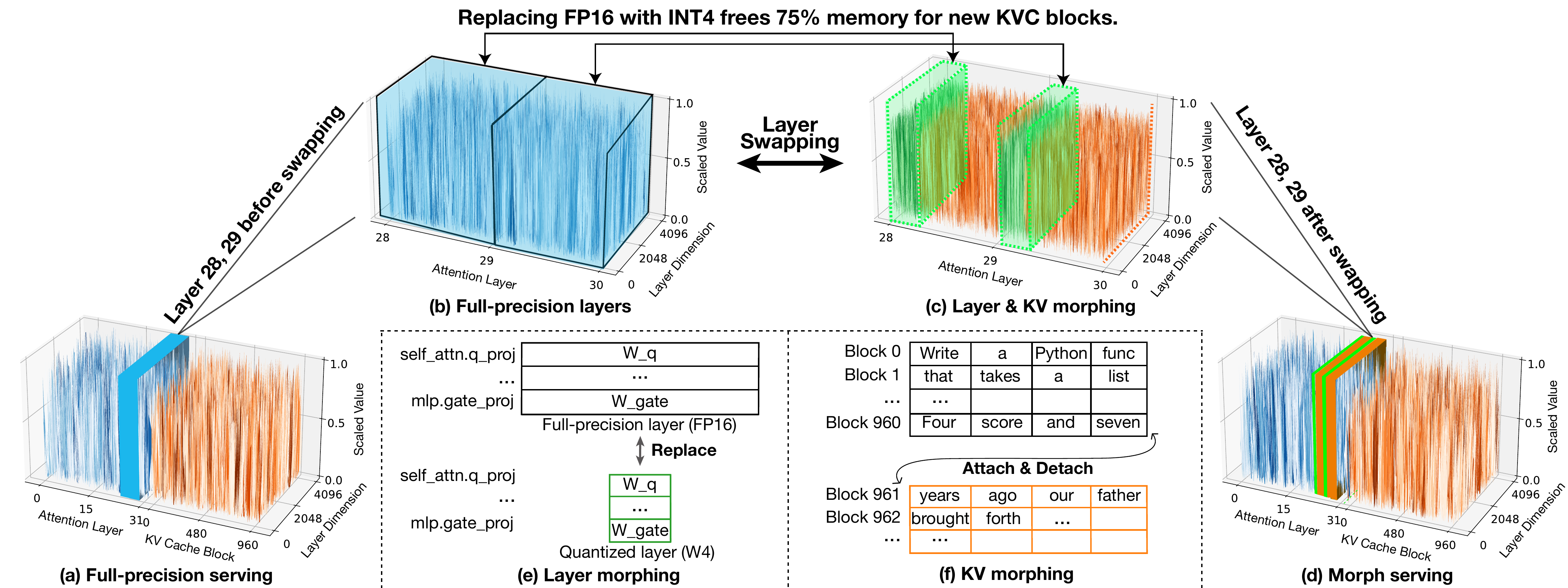}
    \caption{\textbf{Synergy of dynamic layer swapping and elastic KVC resizing.  }
    Figure~(a)–(d) illustrate the model state morphing process: starting from full-precision serving (a), selected layers (b) are replaced with quantized versions (c) without disrupting the inference computation. This process leads to 
    mixed-precision layer serving (d). 
    Figure~(e) shows the detailed decoder layer swapping mechanism. 
    Figure~(f) demonstrates KVC block management under {\kvc}, where newly vacant memory blocks are dynamically reallocated to KVC or deallocated from KVC based on real-time workload shifts. 
    {\kvc} reduces the request preemption rate for decoding and incoming request queueing time for prefilling. 
    }
    \label{fig:layer_morphing}
    \vspace{-12pt} 
\end{figure*}

To enable efficient and non-disruptive layer replacement during inference, {\system} leverages the precomputed layer swapping sequence from offline profiling to guide the dynamic runtime adaptation mechanism. This mechanism consists of two key components: (1)~model preloading with kernel precompilation, which ensures that both full-precision and quantized versions of layers are memory-resident and ready for execution; 
and (2)~asynchronous layer swapping, which allows selected layers to be swapped between CPU and GPU memory on-the-fly without blocking inference.  

\begin{itemize}[leftmargin=*]
    \item \textit{Model Preloading and Kernel Precompilation.}
    Prior to serving, all decoder layer variants (e.g., FP16, INT8, INT4, and INT3)\footnote{{``INT\#'' denotes \textit{weight-only quantization} (W\#). 
    }}
    are preloaded into a contiguous, pinned CPU memory region, while the full-precision model replica is loaded into a preallocated contiguous GPU memory, as shown in the \hyperref[fig:system_design]{Figure~\ref{fig:system_design}}.
    {\system} tracks the memory addresses of all layer variants, enabling efficient direct memory copies for layer swapping. 
    To avoid runtime latency, inference kernels corresponding to precision levels are precompiled in advance. 
    We also implement kernel fusion to optimize performance, while the rest of the serving pipeline reuses state-of-the-art techniques---such as {\PagedAttention}~\cite{kwon2023efficient_vllm} and {\FlashAttention}~\cite{dao2022flashattention, dao2023flashattention_v2}---to ensure compatibility and efficiency.

    \item \textit{Asynchronous In-place Layer Swapping.}  
    At runtime, {\system} performs in-place layer swapping using asynchronous CUDA streams to avoid interference with ongoing decoding. As illustrated in \hyperref[fig:layer_morphing]{Figure~\ref{fig:layer_morphing}}, when layers 28 and 29 are selected for replacement, the swapping process is launched asynchronously while earlier layers (e.g., 0–27) continue computation without interruption. Full-precision layers are safely discarded from GPU memory since their backup copies reside in pinned CPU memory, and quantized variants are copied into the same memory addresses to avoid pointer remapping. Due to the relatively compact size of each decoder layer (e.g., ~0.4 GB for FP16 and ~0.1 GB for INT4 in Llama 2 7B), the PCIe transfer latency is minimal - approximately 4 ms for INT4 and 16 ms for FP16 for Llama2 7B on PCIe Gen4 with up to 26-28 GB/s bandwidth. In practice, the complete layer swapping process for an INT4 variant—including memory transfer and reconstruction—takes approximately 6 ms and is fully overlapped with decoding, resulting in negligible TPOT overhead. Additional performance breakdowns are provided in \hyperref[sec:experiment]{Section~\ref{sec:experiment}}.
\end{itemize}

\begin{figure*}[t!]
    \centering
    \includegraphics[width=0.88\textwidth]{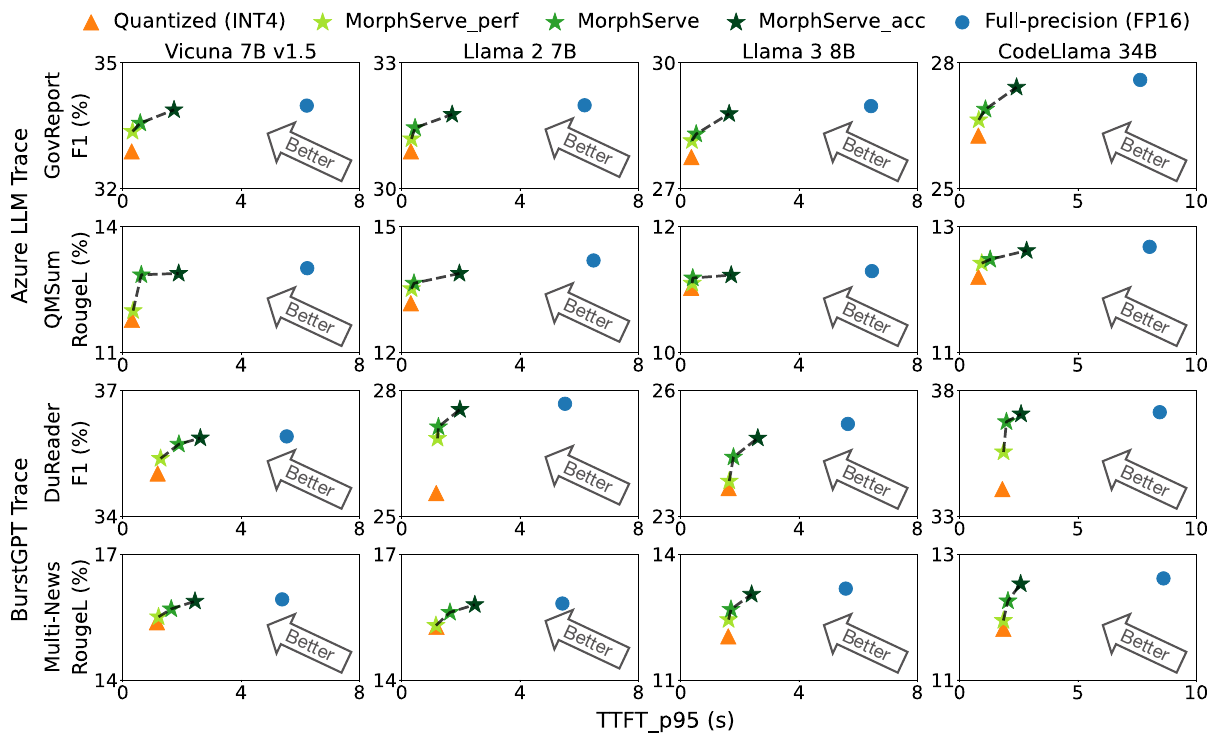}
    \vspace{-0.5em} 
    \caption{
    \textbf{{MorphServe} provides the best latency–accuracy tradeoff across four models and two traces, with four datasets.}
        {MorphServe} in accuracy mode (dark green) reduces P95 TTFT by $2.2\times$–$3.9\times$ compared to full-precision serving while maintaining comparable generation quality. In performance mode (light green), {\system} consistently outperforms INT4 quantized models in output quality with no additional latency overhead.}
    \label{fig:ttft_acc}
    \vspace{-1.5em} 
\end{figure*}

\subsection{{\kvc}: Elastic KVC Resizing}
\label{sec:kv_morphing}

To support bursty workloads and fluctuating memory demands, {\system} integrates dynamic layer swapping with {\kvc}, a mechanism for elastic resizing of KVC blocks.
{\kvc does not quantize or compress existing KV caches.
It reallocates GPU memory freed by weight quantization (e.g., FP16$\!\rightarrow$W4)
to elastically expand the KV cache capacity during high load.}
This section addresses two key questions:
(1) how {\kvc} dynamically allocates and releases KVC blocks in response to runtime memory pressure, and
(2) how it collaborates with layer swapping to maintain serving efficiency under peak load.

{\kvc} is triggered when the Serving Monitor detects insufficient GPU memory to allocate KVC blocks for incoming request \emph{prefilling} or ongoing \emph{decoding}. To free memory, {\system} initiates layer swapping, replacing selected full-precision layers with quantized variants. This reduces the model’s memory footprint—e.g., replacing an FP16 layer with INT4 can save up to \emph{75\%} memory, as shown in \hyperref[fig:layer_morphing]{Figure~\ref{fig:layer_morphing}}—enabling allocation of new KVC blocks.

{\kvc} extends {\PagedAttention}~\cite{kwon2023efficient_vllm} with kernel-level support for \emph{on-demand KVC block allocation/deallocation}, implemented through memory mapping without requiring kernel recompilation. All resizing operations are executed asynchronously using separate CUDA streams to avoid interference with ongoing decoding.

Unlike static preallocation strategies (e.g., in vLLM~\cite{kwon2023efficient_vllm}), {\kvc} adjusts KVC capacity dynamically based on real-time memory availability. Once the pressure subsides, both temporary KVC blocks and quantized layers are released and restored to their full-precision state, ensuring memory reuse and accuracy recovery.

As a result, {\kvc} enhances system efficiency across both the prefilling and decoding phases under high-load conditions.

\begin{itemize}[leftmargin=*]
    \item \textit{Reducing Queueing Time and TTFT During Prefilling.}  
    Under static scheduling, incoming requests may queue indefinitely when no GPU memory is available for KV allocation. Since FIFO schedulers typically release memory only after a request finishes decoding, long queueing delays directly translate into TTFT violations. In {\system}, {\kvc} is triggered when the queue length or wait time exceeds a threshold, proactively attaching new KV blocks to admit pending requests. This significantly reduces queueing time and improves TTFT under bursty traffic.

    \item \textit{Reducing Preemption and Improving TPOT During Decoding.}  
    In the decoding phase, requests are preempted if no KV blocks are available, forcing swaps to host memory or full recomputation, both of which introduce delays and degrade TPOT and end-to-end latency. By dynamically attaching KV blocks at runtime, {\system} reduces preemption events and maintains decoding continuity, leading to better overall system responsiveness.
\end{itemize}\vspace{-10pt} 

Together, these improvements enable {\system} to utilize GPU memory more efficiently across load conditions, mitigate bottlenecks under saturation, and achieve a balanced trade-off between accuracy and responsiveness in volatile serving scenarios.



\section{Experiment}
\label{sec:experiment}

\begin{table*}[t]
\caption{Impact of mixed-precision serving on Llama 3 8B (BookSum Chapters dataset, 6K input / 2K output tokens).}\label{tab:mixed_precision_long}
\centering
\resizebox{0.9\textwidth}{!}{%
\begin{tabular}{|c|c|c|c|}
\hline
\textbf{Generation Strategy} & \textbf{F1 Score} & \textbf{ROUGE-L} & \textbf{Within Accuracy Bound} \\ \hline
All tokens with INT4 (W4) & 14.4716 & 12.1598 & \textbf{Lower Bound} \\ \hline
First 1K tokens FP16, Last 1K tokens W4 & 16.6112 & 15.2346 & \textbf{Yes} \\ \hline
First 1K tokens W4, Last 1K tokens FP16 & 14.7101 & 12.3213 & \textbf{Yes} \\ \hline
First 512 FP16, Middle 1K W4, Last 512 FP16 & 16.0802 & 14.4013 & \textbf{Yes} \\ \hline
First 512 W4, Middle 1K FP16, Last 512 W4 & 16.0413 & 13.1127 & \textbf{Yes} \\ \hline
Switch precision every 256 tokens, starting with W4 & 15.2634 & 13.0353 & \textbf{Yes} \\ \hline
Switch precision every 256 tokens, starting with FP16 & 16.0012 & 14.0146 & \textbf{Yes} \\ \hline
All tokens with full precision (FP16) & \textbf{17.6458} & \textbf{15.4847} & \textbf{Upper Bound} \\ \hline
\end{tabular}
}
\end{table*}

\begin{figure*}[t]
    \centering
    \includegraphics[width=0.94\textwidth]{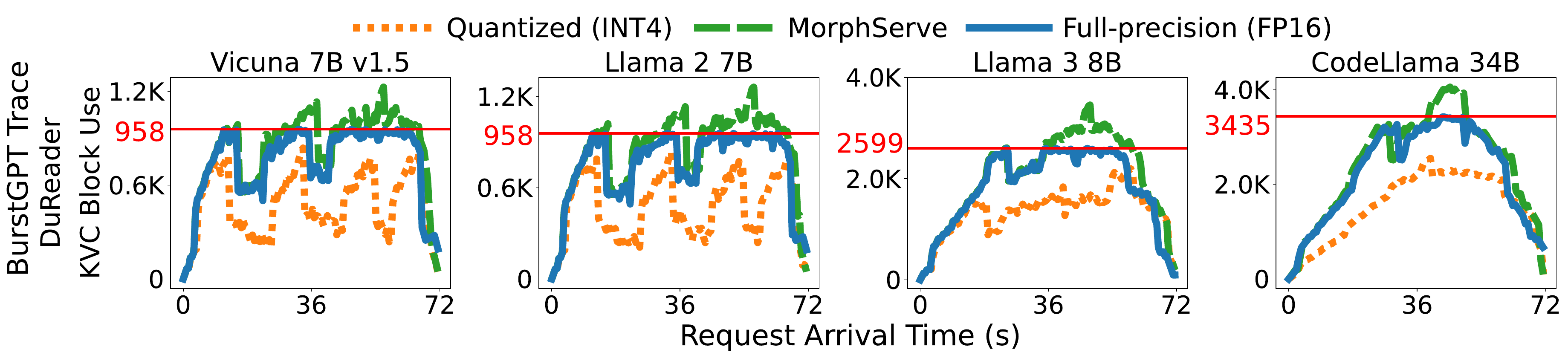}
    \caption{\textbf{MorphServe dynamically adapts KVC block capacity to workload fluctuations. }
   The red line indicates the KV cache capacity under full-precision serving. {\system} (green) elastically attaches new KV blocks during peak loads, pushing the saturation boundary and preventing request preemption or KVC swapping in the full-precision baseline (blue). Static quantization (orange) underutilizes memory due to its fixed configuration, even when resource headroom is available.}
    \label{fig:kvc_adaptation}
\end{figure*}

\begin{figure*}[t]
    \centering
    \includegraphics[width=0.94\textwidth]{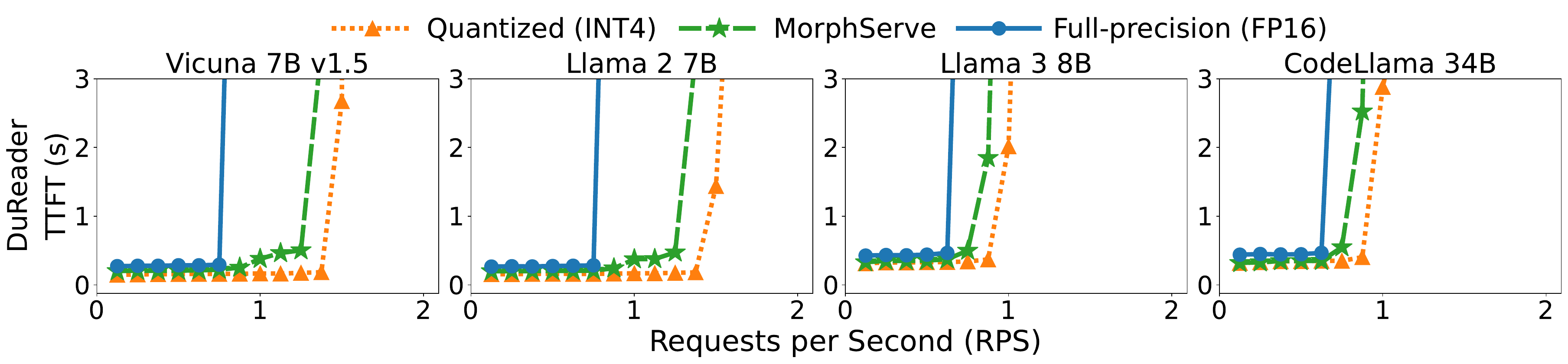}
    \caption{{\system} delays saturation and achieves up to $1.83\times$ throughput than full-precision serving under increasing request rates.}  
    \label{fig:rps_ttft}
\end{figure*}

\begin{figure*}[t]
    \centering
    \includegraphics[width=0.94\textwidth]{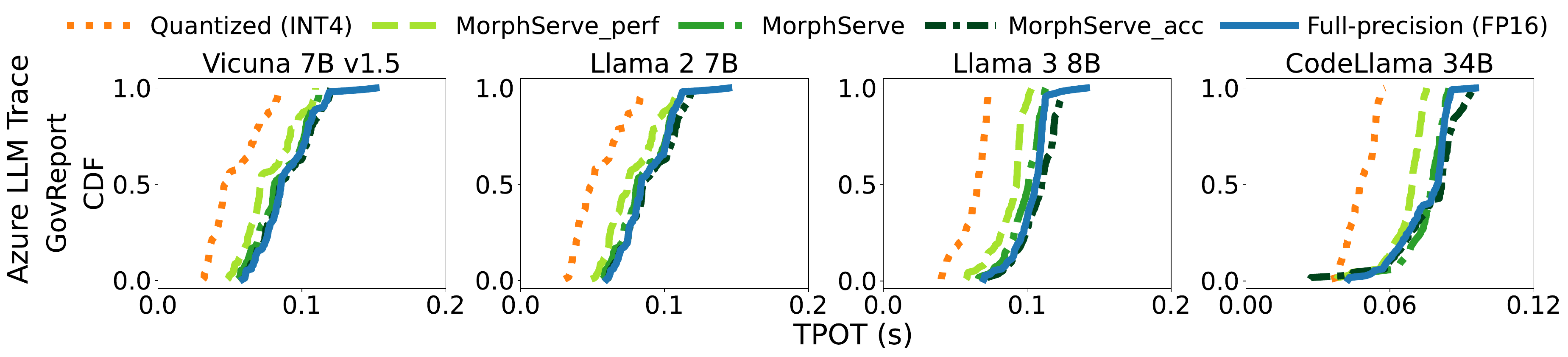}
    \caption{\textbf{{\system} incurs negligible runtime overhead while improving tail TPOT latency.}
        MorphServe (green) achieves comparable average TPOT latency to the full-precision baseline (blue), while reducing P99 latency by up to $1.23\times$. Performance mode (light green) improves the average TPOT by up to $1.17\times$ through aggressive layer morphing.
    }  
    \label{fig:tpot_cdf}
\end{figure*}

\begin{table*}[h]
\caption{\textbf{Performance of MorphServe with AWQ and uniform INT4 quantization vs. FP16 on Llama 2 7B, 
using the DuReader dataset and BurstGPT trace.} MorphServe improves accuracy over static
quantization, reducing accuracy degradation by 83.57\% for AWQ, 74.67\% for Uniform, while
lowering P95 TTFT by over 77\% compared to FP16.}\label{tab:perf_ablation_llama2}
\centering
\begin{tabular}{|c|c|c|c|c|}
\hline
Quantization (INT4) & Metric & Static Quantization & {\system} & Full Precision \\
\hline
& TTFT P95 (s) & 1.1686 & 1.2420 & 5.5223 \\
\cline{2-5}
AWQ & TPOT P99 (s) & 0.0735 & 0.1064 & 0.1241 \\
\cline{2-5}
& F1 Score & 25.55 & 27.33 & 27.68 \\
\hline
& TTFT P95 (s) & 1.1321 & 1.2144 & 5.5223 \\
\cline{2-5}
Uniform & TPOT P99 (s) & 0.0728 & 0.1053 & 0.1241 \\
\cline{2-5}
& F1 Score & 24.87 & 26.94 & 27.68 \\
\hline
\end{tabular}
\end{table*}

\begin{table*}[h]
\caption{Cross-dataset evaluation of LIS profiles---perplexity on C4 using LIS calibrated on WikiText-2.}\label{tab:cross_dataset}
\centering
\begin{tabular}{|c|c|c|c|c|c|c|}
\hline
\textbf{Method} & \textbf{1} & \textbf{2} & \textbf{4} & \textbf{8} & \textbf{16} & \textbf{32 (Fully Quantized)} \\ \hline
Front-to-Back & \textbf{7.0793} & 7.0822 & 7.0933 & 7.1119 & 7.1524 & \textbf{7.2463} \\ \hline
Back-to-Front & 7.0871 & 7.0938 & 7.1031 & 7.1224 & 7.1712 & \textbf{7.2463} \\ \hline
Random & 7.0871 & 7.0913 & 7.1012 & 7.1234 & 7.1676 & \textbf{7.2463} \\ \hline
LIS & 7.0801 & \textbf{7.0818} & \textbf{7.0912} & \textbf{7.1107} & \textbf{7.1497} & \textbf{7.2463} \\ \hline
\end{tabular}
\end{table*}

\textbf{Evaluation Setup}.
We evaluate {\system} across a diverse range of LLM architectures, workload traces, and tasks. 
We consider four representative models: Vicuna 7B v1.5~\cite{vicuna_7b_v1.5}, Llama 2 7B~\cite{touvron2023llama2}, Llama 3 8B~\cite{grattafiori2024llama3}, and CodeLlama 34B~\cite{roziere2023codellama}, spanning multiple scales and attention types---including Multi-Head Attention (MHA)~\cite{vaswani2017attention} and Grouped-Query Attention (GQA)~\cite{ainslie2023gqa}.
We test two real-world LLM inference workload traces: the BurstGPT trace~\cite{wang2024burstgpt} and the Azure LLM Inference trace~\cite{stojkovic2025dynamollm, azure_LLM_inference_trace}. 
We report results from a representative 72-second trace snippet (\hyperref[fig:motivations]{Figure~\ref{fig:motivations}}) for both workloads, though {\system} is effective across the full traces.
The request arrival rates
of each trace are downscaled by 1.75$\times$ and 4.75$\times$ to fit our hardware environment. 
To evaluate generation quality, 
we use four public datasets: GovReport~\cite{gov_report} and Multi-News~\cite{fabbri2019multi} (long-form summarization), QMSum~\cite{zhong2021qmsum} (query-based summarization), and DuReader~\cite{he2017dureader} (reading comprehension). 
For each test, we align workload timestamps with context passages from the datasets. Prompt and response lengths are set to \emph{512} and \emph{256} tokens for Vicuna 7B v1.5 and Llama 2 7B, and to \emph{1024} and \emph{512} tokens for Llama 3 8B and CodeLlama 34B. We report F1 and Rouge-L scores to assess generation quality. End-to-end experiments for Vicuna 7B v1.5, Llama 2 7B, and Llama 3 8B are conducted on an NVIDIA L4 GPU with 24 GB HBM and 256 GB of CPU DRAM, while CodeLlama 34B is evaluated on an A100 server with 80 GB HBM and 2 TB of CPU DRAM.

\textbf{Implementation.}
{\system} is implemented on top of SwiftLLM~\cite{jiang2024neo_swiftLLM, swiftllm_github_repo}, a lightweight and modular LLM inference framework that reproduces vLLM~\cite{kwon2023efficient_vllm} performance with simplified components. 
We added approximately 2,200 lines of Python and 500 lines of C++/CUDA to support {\system}’s optimized KVC management and attention kernel extensions, which enable efficient layer swapping and KVC resizing at runtime.  

\textbf{Comparison Scope and Baselines.} LLM serving frameworks differ widely in objectives: static systems like vLLM~\cite{kwon2023efficient_vllm}, Sarathi~\cite{agrawal2023sarathi}, and FastServe~\cite {wu2023fastserve} focus on fixed-precision serving, while speculative or input-aware approaches like MARLIN~\cite{frantar2025marlin_mixed_precision}, LayerSkip~\cite{Elhoushi_2024}, and FlexiDepth~\cite{luo2025adaptive} optimize inference paths offline. In contrast, MorphServe adapts in real time, dynamically navigating the performance–accuracy tradeoff during live serving. We benchmark against two representative baselines: FP16 (upper-bound accuracy) and INT4 (performance-optimized static quantization). MorphServe is evaluated in two runtime modes: \textbf{accuracy mode}, which conservatively triggers layer swapping to preserve quality, and \textbf{performance mode}, which aggressively swaps layers to improve throughput under memory pressure. All setups use the same inference engine, and AWQ~\cite{lin2024awq} is adopted for its efficient kernel support, though MorphServe is quantization-method agnostic.

\subsection{Main Results}

\textbf{TTFT and Accuracy.}
As shown in \hyperref[fig:ttft_acc]{Figure~\ref{fig:ttft_acc}}, {\system} significantly reduces P95 TTFT latency (95th-percentile TTFT latency) while preserving output quality in all model-trace-dataset configurations. Compared to full-precision baselines, {\system} reduces the P95 TTFT by $2.9\times$–$15.7\times$ ($2.2\times$–$3.9\times$ in accuracy mode and $3.4\times$–$19.5\times$ in performance mode) while maintaining quality within $0.51\%$–$3.82\%$ degradation on F1 or Rouge-L scores, as low as $0.11\%$–$2.18\%$ in accuracy mode.
In contrast, static quantization exhibits $2.34\%$-$9.47\%$ degradation compared to full-precision inference, due to suffering from persistent quality loss across the entire serving lifetime. In particular, {\system} excels in long-context datasets such as GovReport, leveraging {\quant} and {\kvc} to optimize memory and computing efficiency. {\system} with different configurations (green stars) visualizes the ability to navigate the latency-accuracy Pareto frontier, offering the best balance of performance and quality based on real-time workload shifts.

\textbf{Generation Quality under Mixed-Precision Serving.}
MorphServe adopts a conservative adaptation strategy, replacing full-precision layers with high-quality quantized counterparts (e.g., AWQ) rather than skipping them at the token level~\cite{dynamic-layer-selection}. This design preserves stable and bounded generation quality during dynamic adaptation. As shown in \hyperref[tab:mixed_precision_long]{Table~\ref{tab:mixed_precision_long}}, even for long-context tasks (BookSum Chapters, 6K input and 2K output tokens), all mixed-precision configurations maintain performance between the fully quantized and full-precision bounds, ensuring controllable quality under frequent precision switching. Additional perplexity evaluations across models are provided in \hyperref[appendix:eval]{Appendix}. A verbatim example of MorphServe’s output under mixed-precision serving is also in the \hyperref[appendix:eval]{Appendix}.

\textbf{Workload Adaptation and Saturation Resilience.}
As shown in \hyperref[fig:kvc_adaptation]{Figure~\ref{fig:kvc_adaptation}}, {\system} adaptively manages KVC block capacity in response to fluctuating load. In the full-precision baseline, KVC usage saturates the static capacity during peak periods, resulting in elevated queueing delays, request preemption, and frequent KVC swapping, which can lead to SLO violations. Static quantization, while reducing the memory footprint, 
degrades model accuracy and underutilizes GPU memory, even during low-load periods. 
{\system} attaches new blocks during bursty traffic and releases them as load subsides, enabled by the synergistic {\quant} and {\kvc} mechanism. 
{\system} improves overall KVC memory utilization and output accuracy by $29.29\%$ and $3.58\%$, respectively, compared to static quantization. The adaptation allows {\system} to expand KVC usage by up to $32.97\%$ beyond the full-precision limit when needed, and reduce the queueing delay by up to $3.8\times$. 
{\system} also mitigates request preemption and KVC swapping under saturation conditions. 
This enhances system responsiveness and improves token-level efficiency, contributing to reduced end-to-end request latency.

\textbf{Throughput.} 
In \hyperref[fig:rps_ttft]{Figure~\ref{fig:rps_ttft}}, we compare {\system} with baselines on DuReader under varying request rates. All configurations maintain low TTFT at low RPS, but as load increases, full-precision inference encounters the threshold, where TTFT spikes abruptly due to memory exhaustion and queueing delays. In contrast, {\system} consistently pushes back this saturation point, achieving $1.6\times$–$1.83\times$ higher throughput than full-precision serving across all evaluated models.


\textbf{TPOT Tail Latency.}
\hyperref[fig:tpot_cdf]{Figure~\ref{fig:tpot_cdf}} presents the cumulative distribution (CDF) of
time-per-output-token (TPOT) across two datasets and four models under the Azure LLM trace.
{\system} delivers average TPOT comparable to full-precision serving while improving
P95 and P99 tail latency by up to $1.06\times$ and $1.23\times$, respectively.
These gains arise from eliminating preemption stalls and avoiding KVC swapping or recomputation,
the main sources of long-tail delays.
In performance mode, TPOT decreases by $1.11$-$1.17\times$,
whereas accuracy mode adds up to $1.06\times$ overhead due to preserving more FP16 layers
and applying stricter layer-swapping thresholds, occasionally causing short queuing under load.
The improvement results from faster inference on quantized layers and highly efficient
swapping kernels, 
confirming that {\system} introduces negligible runtime overhead while effectively reducing tail latency.


\subsection{Ablation Study}
\textbf{Effectiveness Across Quantization Algorithms.}
{{\system} is agnostic to specific quantization schemes.
We evaluate it under both AWQ and uniform INT4 using identical datasets and traces.
As shown in \hyperref[tab:perf_ablation_llama2]{Table~\ref{tab:perf_ablation_llama2}} (and Table 5 in the appendix),
{\system} consistently improves F1 and Rouge-L over static quantization
while maintaining significant latency advantages over full-precision serving.
These results confirm that its gains arise from runtime adaptation
rather than the choice of quantization algorithm.}

\textbf{Generalization of Layer Importance Score.}
{\hyperref[tab:cross_dataset]{Table~\ref{tab:cross_dataset}} shows the generalization of LIS profiles, which consistently outperforms front-to-back, back-to-front, and random layer orders on C4 with calibration on Wiki2,
indicating that LIS captures intrinsic layer importance and generalizes across datasets without re-profiling.}

\section{Discussion and Future Work}

\noindent\textbf{User Tolerance.}
{Since fully quantized models (e.g., AWQ) are widely acceptable, 
{\system}'s outputs remain within user tolerance. 
It maintains full-precision decoding under light load and bounds quality variation 
under adaptation, achieving stable perceptual quality with adaptive efficiency.}

\noindent\textbf{Scalability.}
{{\system} scales naturally to distributed serving with data, tensor, pipeline, 
and sequence parallelism. Each GPU runs an independent executor that swaps layers 
and resizes KV caches locally.}

\noindent\textbf{Fairness.}
The continuous batching~\cite{yu2022orca} preserves FIFO ordering, 
while chunked prefill~\cite{agrawal2023sarathi} and fine-grained KV management~\cite{kwon2023efficient_vllm}
prevent starvation. 
Combined with these, MorphServe serving ensures fair and stable serving under bursty conditions.

\noindent\textbf{Future Work.}
{{\system} can extend beyond weight-only quantization to include activation and 
KV quantization. Such extensions promise higher efficiency but raise calibration 
and scheduling challenges. We plan to explore these for edge deployment and 
robust, energy-efficient adaptation.}

\section{Conclusion}
\label{sec:conclusion}

This paper presents {\system}, a novel workload-aware LLM serving framework based on morphological adaptation. Guided by a unified Layer Importance Score, {\system} dynamically adjusts model precision through {\quant} and KV cache memory capacity through {\kvc}, in a coordinated manner based on real-time resource usage.  
{\system} maintains high-quality inference under normal conditions and adapts gracefully during overload periods. 
Extensive experiments show that {\system} improves SLO compliance, memory efficiency, and robustness, while maintaining high generation quality along the efficiency–accuracy Pareto frontier.


\newpage
\bibliography{reference}

@inproceedings{kwon2023efficient_vllm,
  title={Efficient memory management for large language model serving with pagedattention},
  author={Kwon, Woosuk and Li, Zhuohan and Zhuang, Siyuan and Sheng, Ying and Zheng, Lianmin and Yu, Cody Hao and Gonzalez, Joseph and Zhang, Hao and Stoica, Ion},
  booktitle={Proceedings of the 29th Symposium on Operating Systems Principles},
  pages={611--626},
  year={2023}
}

@article{dao2022flashattention,
  title={Flashattention: Fast and memory-efficient exact attention with io-awareness},
  author={Dao, Tri and Fu, Dan and Ermon, Stefano and Rudra, Atri and R{\'e}, Christopher},
  journal={Advances in neural information processing systems},
  volume={35},
  pages={16344--16359},
  year={2022}
}

@article{dao2023flashattention_v2,
  title={Flashattention-2: Faster attention with better parallelism and work partitioning},
  author={Dao, Tri},
  journal={arXiv preprint arXiv:2307.08691},
  year={2023}
}

@article{lin2024awq,
  title={Awq: Activation-aware weight quantization for on-device llm compression and acceleration},
  author={Lin, Ji and Tang, Jiaming and Tang, Haotian and Yang, Shang and Chen, Wei-Ming and Wang, Wei-Chen and Xiao, Guangxuan and Dang, Xingyu and Gan, Chuang and Han, Song},
  journal={Proceedings of Machine Learning and Systems},
  volume={6},
  pages={87--100},
  year={2024}
}

@article{lin2024duquant,
  title={Duquant: Distributing outliers via dual transformation makes stronger quantized llms},
  author={Lin, Haokun and Xu, Haobo and Wu, Yichen and Cui, Jingzhi and Zhang, Yingtao and Mou, Linzhan and Song, Linqi and Sun, Zhenan and Wei, Ying},
  journal={Advances in Neural Information Processing Systems},
  volume={37},
  pages={87766--87800},
  year={2024}
}

@inproceedings{xiao2023smoothquant,
  title={Smoothquant: Accurate and efficient post-training quantization for large language models},
  author={Xiao, Guangxuan and Lin, Ji and Seznec, Mickael and Wu, Hao and Demouth, Julien and Han, Song},
  booktitle={International Conference on Machine Learning},
  pages={38087--38099},
  year={2023},
  organization={PMLR}
}

@article{yao2022zeroquant,
  title={Zeroquant: Efficient and affordable post-training quantization for large-scale transformers},
  author={Yao, Zhewei and Yazdani Aminabadi, Reza and Zhang, Minjia and Wu, Xiaoxia and Li, Conglong and He, Yuxiong},
  journal={Advances in Neural Information Processing Systems},
  volume={35},
  pages={27168--27183},
  year={2022}
}

@inproceedings{nagel2020up_adaround,
  title={Up or down? adaptive rounding for post-training quantization},
  author={Nagel, Markus and Amjad, Rana Ali and Van Baalen, Mart and Louizos, Christos and Blankevoort, Tijmen},
  booktitle={International conference on machine learning},
  pages={7197--7206},
  year={2020},
  organization={PMLR}
}

@article{lin2024qserve,
  title={Qserve: W4a8kv4 quantization and system co-design for efficient llm serving},
  author={Lin, Yujun and Tang, Haotian and Yang, Shang and Zhang, Zhekai and Xiao, Guangxuan and Gan, Chuang and Han, Song},
  journal={arXiv preprint arXiv:2405.04532},
  year={2024}
}

@article{dettmers2023spqr,
  title={Spqr: A sparse-quantized representation for near-lossless llm weight compression},
  author={Dettmers, Tim and Svirschevski, Ruslan and Egiazarian, Vage and Kuznedelev, Denis and Frantar, Elias and Ashkboos, Saleh and Borzunov, Alexander and Hoefler, Torsten and Alistarh, Dan},
  journal={arXiv preprint arXiv:2306.03078},
  year={2023}
}

@article{kim2023squeezellm,
  title={Squeezellm: Dense-and-sparse quantization},
  author={Kim, Sehoon and Hooper, Coleman and Gholami, Amir and Dong, Zhen and Li, Xiuyu and Shen, Sheng and Mahoney, Michael W and Keutzer, Kurt},
  journal={arXiv preprint arXiv:2306.07629},
  year={2023}
}

@inproceedings{li2024norm_quant,
  title={Norm tweaking: High-performance low-bit quantization of large language models},
  author={Li, Liang and Li, Qingyuan and Zhang, Bo and Chu, Xiangxiang},
  booktitle={Proceedings of the AAAI Conference on Artificial Intelligence},
  volume={38},
  number={17},
  pages={18536--18544},
  year={2024}
}

@inproceedings{wang2020towards_bitsplit,
  title={Towards accurate post-training network quantization via bit-split and stitching},
  author={Wang, Peisong and Chen, Qiang and He, Xiangyu and Cheng, Jian},
  booktitle={International Conference on Machine Learning},
  pages={9847--9856},
  year={2020},
  organization={PMLR}
}

@article{frantar2022gptq,
  title={Gptq: Accurate post-training quantization for generative pre-trained transformers},
  author={Frantar, Elias and Ashkboos, Saleh and Hoefler, Torsten and Alistarh, Dan},
  journal={arXiv preprint arXiv:2210.17323},
  year={2022}
}

@article{zhao2024atom,
  title={Atom: Low-bit quantization for efficient and accurate llm serving},
  author={Zhao, Yilong and Lin, Chien-Yu and Zhu, Kan and Ye, Zihao and Chen, Lequn and Zheng, Size and Ceze, Luis and Krishnamurthy, Arvind and Chen, Tianqi and Kasikci, Baris},
  journal={Proceedings of Machine Learning and Systems},
  volume={6},
  pages={196--209},
  year={2024}
}

@article{wang2024burstgpt,
  title={BurstGPT: A Real-world Workload Dataset to Optimize LLM Serving Systems},
  author={Wang, Yuxin and Chen, Yuhan and Li, Zeyu and Kang, Xueze and Tang, Zhenheng and He, Xin and Guo, Rui and Wang, Xin and Wang, Qiang and Zhou, Amelie Chi and others},
  journal={arXiv preprint arXiv:2401.17644},
  year={2024}
}

@inproceedings{he2024deferred_llm_batching,
  title={Deferred continuous batching in resource-efficient large language model serving},
  author={He, Yongjun and Lu, Yao and Alonso, Gustavo},
  booktitle={Proceedings of the 4th Workshop on Machine Learning and Systems},
  pages={98--106},
  year={2024}
}

@article{model_compression_2018_iclr,
  title={Model compression via distillation and quantization},
  author={Polino, Antonio and Pascanu, Razvan and Alistarh, Dan},
  journal={arXiv preprint arXiv:1802.05668},
  year={2018}
}

@inproceedings{sun2024llumnix,
  title={Llumnix: Dynamic scheduling for large language model serving},
  author={Sun, Biao and Huang, Ziming and Zhao, Hanyu and Xiao, Wencong and Zhang, Xinyi and Li, Yong and Lin, Wei},
  booktitle={18th USENIX Symposium on Operating Systems Design and Implementation (OSDI 24)},
  pages={173--191},
  year={2024}
}

@article{shao2023omniquant,
  title={Omniquant: Omnidirectionally calibrated quantization for large language models},
  author={Shao, Wenqi and Chen, Mengzhao and Zhang, Zhaoyang and Xu, Peng and Zhao, Lirui and Li, Zhiqian and Zhang, Kaipeng and Gao, Peng and Qiao, Yu and Luo, Ping},
  journal={arXiv preprint arXiv:2308.13137},
  year={2023}
}

@article{hu2022lora,
  title={Lora: Low-rank adaptation of large language models.},
  author={Hu, Edward J and Shen, Yelong and Wallis, Phillip and Allen-Zhu, Zeyuan and Li, Yuanzhi and Wang, Shean and Wang, Lu and Chen, Weizhu and others},
  journal={ICLR},
  volume={1},
  number={2},
  pages={3},
  year={2022}
}

@inproceedings{wu2024dlora,
  title={$\{$dLoRA$\}$: Dynamically orchestrating requests and adapters for $\{$LoRA$\}$$\{$LLM$\}$ serving},
  author={Wu, Bingyang and Zhu, Ruidong and Zhang, Zili and Sun, Peng and Liu, Xuanzhe and Jin, Xin},
  booktitle={18th USENIX Symposium on Operating Systems Design and Implementation (OSDI 24)},
  pages={911--927},
  year={2024}
}

@article{gao2024disp_pruning,
  title={Disp-llm: Dimension-independent structural pruning for large language models},
  author={Gao, Shangqian and Lin, Chi-Heng and Hua, Ting and Tang, Zheng and Shen, Yilin and Jin, Hongxia and Hsu, Yen-Chang},
  journal={Advances in Neural Information Processing Systems},
  volume={37},
  pages={72219--72244},
  year={2024}
}

@article{sreenivas2024llm_pruning,
  title={Llm pruning and distillation in practice: The minitron approach},
  author={Sreenivas, Sharath Turuvekere and Muralidharan, Saurav and Joshi, Raviraj and Chochowski, Marcin and Mahabaleshwarkar, Ameya Sunil and Shen, Gerald and Zeng, Jiaqi and Chen, Zijia and Suhara, Yoshi and Diao, Shizhe and others},
  journal={arXiv preprint arXiv:2408.11796},
  year={2024}
}

@article{ma2023llm_pruning,
  title={Llm-pruner: On the structural pruning of large language models},
  author={Ma, Xinyin and Fang, Gongfan and Wang, Xinchao},
  journal={Advances in neural information processing systems},
  volume={36},
  pages={21702--21720},
  year={2023}
}

@inproceedings{yu2022orca,
  title={Orca: A distributed serving system for $\{$Transformer-Based$\}$ generative models},
  author={Yu, Gyeong-In and Jeong, Joo Seong and Kim, Geon-Woo and Kim, Soojeong and Chun, Byung-Gon},
  booktitle={16th USENIX Symposium on Operating Systems Design and Implementation (OSDI 22)},
  pages={521--538},
  year={2022}
}

@inproceedings{Elhoushi_2024,
   title={LayerSkip: Enabling Early Exit Inference and Self-Speculative Decoding},
   url={http://dx.doi.org/10.18653/v1/2024.acl-long.681},
   DOI={10.18653/v1/2024.acl-long.681},
   booktitle={Proceedings of the 62nd Annual Meeting of the Association for Computational Linguistics (Volume 1: Long Papers)},
   publisher={Association for Computational Linguistics},
   author={Elhoushi, Mostafa and Shrivastava, Akshat and Liskovich, Diana and Hosmer, Basil and Wasti, Bram and Lai, Liangzhen and Mahmoud, Anas and Acun, Bilge and Agarwal, Saurabh and Roman, Ahmed and Aly, Ahmed and Chen, Beidi and Wu, Carole-Jean},
   year={2024},
   pages={12622–12642} }

@misc{luo2025adaptive,
      title={Adaptive Layer-skipping in Pre-trained LLMs}, 
      author={Xuan Luo and Weizhi Wang and Xifeng Yan},
      year={2025},
      eprint={2503.23798},
      archivePrefix={arXiv},
      primaryClass={cs.CL},
      url={https://arxiv.org/abs/2503.23798}, 
}

@inproceedings{cai2024edge_llm,
  title={Edge-llm: A collaborative framework for large language model serving in edge computing},
  author={Cai, Fenglong and Yuan, Dong and Yang, Zhe and Cui, Lizhen},
  booktitle={2024 IEEE International Conference on Web Services (ICWS)},
  pages={799--809},
  year={2024},
  organization={IEEE}
}

@inproceedings{fu2024serverlessllm,
  title={$\{$ServerlessLLM$\}$:$\{$Low-Latency$\}$ serverless inference for large language models},
  author={Fu, Yao and Xue, Leyang and Huang, Yeqi and Brabete, Andrei-Octavian and Ustiugov, Dmitrii and Patel, Yuvraj and Mai, Luo},
  booktitle={18th USENIX Symposium on Operating Systems Design and Implementation (OSDI 24)},
  pages={135--153},
  year={2024}
}

@article{jaiswal2025serving_provisioning,
  title={Serving models, fast and slow: optimizing heterogeneous LLM inferencing workloads at scale},
  author={Jaiswal, Shashwat and Jain, Kunal and Simmhan, Yogesh and Parayil, Anjaly and Mallick, Ankur and Wang, Rujia and Amant, Renee St and Bansal, Chetan and R{\"u}hle, Victor and Kulkarni, Anoop and others},
  journal={arXiv preprint arXiv:2502.14617},
  year={2025}
}

@article{cai2024pyramidkv,
  title={Pyramidkv: Dynamic kv cache compression based on pyramidal information funneling},
  author={Cai, Zefan and Zhang, Yichi and Gao, Bofei and Liu, Yuliang and Liu, Tianyu and Lu, Keming and Xiong, Wayne and Dong, Yue and Chang, Baobao and Hu, Junjie and others},
  journal={arXiv preprint arXiv:2406.02069},
  year={2024}
}

@article{zhou2024dynamickv,
  title={DynamicKV: Task-Aware Adaptive KV Cache Compression for Long Context LLMs},
  author={Zhou, Xiabin and Wang, Wenbin and Zeng, Minyan and Guo, Jiaxian and Liu, Xuebo and Shen, Li and Zhang, Min and Ding, Liang},
  journal={arXiv preprint arXiv:2412.14838},
  year={2024}
}

@article{feng2024ada,
  title={Ada-kv: Optimizing kv cache eviction by adaptive budget allocation for efficient llm inference},
  author={Feng, Yuan and Lv, Junlin and Cao, Yukun and Xie, Xike and Zhou, S Kevin},
  journal={arXiv preprint arXiv:2407.11550},
  year={2024}
}

@article{li2024snapkv,
  title={Snapkv: Llm knows what you are looking for before generation},
  author={Li, Yuhong and Huang, Yingbing and Yang, Bowen and Venkitesh, Bharat and Locatelli, Acyr and Ye, Hanchen and Cai, Tianle and Lewis, Patrick and Chen, Deming},
  journal={Advances in Neural Information Processing Systems},
  volume={37},
  pages={22947--22970},
  year={2024}
}

@article{liu2023scissorhands,
  title={Scissorhands: Exploiting the persistence of importance hypothesis for llm kv cache compression at test time},
  author={Liu, Zichang and Desai, Aditya and Liao, Fangshuo and Wang, Weitao and Xie, Victor and Xu, Zhaozhuo and Kyrillidis, Anastasios and Shrivastava, Anshumali},
  journal={Advances in Neural Information Processing Systems},
  volume={36},
  pages={52342--52364},
  year={2023}
}

@article{ainslie2023gqa,
  title={Gqa: Training generalized multi-query transformer models from multi-head checkpoints},
  author={Ainslie, Joshua and Lee-Thorp, James and De Jong, Michiel and Zemlyanskiy, Yury and Lebr{\'o}n, Federico and Sanghai, Sumit},
  journal={arXiv preprint arXiv:2305.13245},
  year={2023}
}

@article{chen2024optimised_gqa,
  title={Optimised grouped-query attention mechanism for transformers},
  author={Chen, Yuang and Zhang, Cheng and Gao, Xitong and Mullins, Robert D and Constantinides, George A and Zhao, Yiren},
  journal={arXiv preprint arXiv:2406.14963},
  year={2024}
}

@article{meng2025transmla,
  title={TransMLA: Multi-head Latent Attention Is All You Need},
  author={Meng, Fanxu and Yao, Zengwei and Zhang, Muhan},
  journal={arXiv preprint arXiv:2502.07864},
  year={2025}
}

@article{chinnakonduru2024weighted_gqa,
  title={Weighted Grouped Query Attention in Transformers},
  author={Chinnakonduru, Sai Sena and Mohapatra, Astarag},
  journal={arXiv preprint arXiv:2407.10855},
  year={2024}
}

@article{liu2024deepseek,
  title={Deepseek-v2: A strong, economical, and efficient mixture-of-experts language model},
  author={Liu, Aixin and Feng, Bei and Wang, Bin and Wang, Bingxuan and Liu, Bo and Zhao, Chenggang and Dengr, Chengqi and Ruan, Chong and Dai, Damai and Guo, Daya and others},
  journal={arXiv preprint arXiv:2405.04434},
  year={2024}
}

@article{touvron2023llama2,
  title={Llama 2: Open foundation and fine-tuned chat models},
  author={Touvron, Hugo and Martin, Louis and Stone, Kevin and Albert, Peter and Almahairi, Amjad and Babaei, Yasmine and Bashlykov, Nikolay and Batra, Soumya and Bhargava, Prajjwal and Bhosale, Shruti and others},
  journal={arXiv preprint arXiv:2307.09288},
  year={2023}
}

@article{grattafiori2024llama3,
  title={The llama 3 herd of models},
  author={Grattafiori, Aaron and Dubey, Abhimanyu and Jauhri, Abhinav and Pandey, Abhinav and Kadian, Abhishek and Al-Dahle, Ahmad and Letman, Aiesha and Mathur, Akhil and Schelten, Alan and Vaughan, Alex and others},
  journal={arXiv preprint arXiv:2407.21783},
  year={2024}
}

@inproceedings{gov_report,
    title = "Efficient Attentions for Long Document Summarization",
    author = "Huang, Luyang  and
      Cao, Shuyang  and
      Parulian, Nikolaus  and
      Ji, Heng  and
      Wang, Lu",
    booktitle = "Proceedings of the 2021 Conference of the North American Chapter of the Association for Computational Linguistics: Human Language Technologies",
    month = jun,
    year = "2021",
    address = "Online",
    publisher = "Association for Computational Linguistics",
    url = "https://aclanthology.org/2021.naacl-main.112",
    doi = "10.18653/v1/2021.naacl-main.112",
    pages = "1419--1436",
}

@article{zhong2021qmsum,
  title={QMSum: A new benchmark for query-based multi-domain meeting summarization},
  author={Zhong, Ming and Yin, Da and Yu, Tao and Zaidi, Ahmad and Mutuma, Mutethia and Jha, Rahul and Awadallah, Ahmed Hassan and Celikyilmaz, Asli and Liu, Yang and Qiu, Xipeng and others},
  journal={arXiv preprint arXiv:2104.05938},
  year={2021}
}

@article{he2017dureader,
  title={Dureader: a chinese machine reading comprehension dataset from real-world applications},
  author={He, Wei and Liu, Kai and Liu, Jing and Lyu, Yajuan and Zhao, Shiqi and Xiao, Xinyan and Liu, Yuan and Wang, Yizhong and Wu, Hua and She, Qiaoqiao and others},
  journal={arXiv preprint arXiv:1711.05073},
  year={2017}
}

@article{fabbri2019multi,
  title={Multi-news: A large-scale multi-document summarization dataset and abstractive hierarchical model},
  author={Fabbri, Alexander R and Li, Irene and She, Tianwei and Li, Suyi and Radev, Dragomir R},
  journal={arXiv preprint arXiv:1906.01749},
  year={2019}
}

@inproceedings{stojkovic2025dynamollm,
  title={Dynamollm: Designing llm inference clusters for performance and energy efficiency},
  author={Stojkovic, Jovan and Zhang, Chaojie and Goiri, {\'I}{\~n}igo and Torrellas, Josep and Choukse, Esha},
  booktitle={2025 IEEE International Symposium on High Performance Computer Architecture (HPCA)},
  pages={1348--1362},
  year={2025},
  organization={IEEE}
}

@article{roziere2023codellama,
  title={Code llama: Open foundation models for code},
  author={Roziere, Baptiste and Gehring, Jonas and Gloeckle, Fabian and Sootla, Sten and Gat, Itai and Tan, Xiaoqing Ellen and Adi, Yossi and Liu, Jingyu and Sauvestre, Romain and Remez, Tal and others},
  journal={arXiv preprint arXiv:2308.12950},
  year={2023}
}

@article{xiong2024layerkv,
  title={Layerkv: Optimizing large language model serving with layer-wise kv cache management},
  author={Xiong, Yi and Wu, Hao and Shao, Changxu and Wang, Ziqing and Zhang, Rui and Guo, Yuhong and Zhao, Junping and Zhang, Ke and Pan, Zhenxuan},
  journal={arXiv preprint arXiv:2410.00428},
  year={2024}
}

@inproceedings{gao2024cost_cachedattention,
  title={$\{$Cost-Efficient$\}$ large language model serving for multi-turn conversations with $\{$CachedAttention$\}$},
  author={Gao, Bin and He, Zhuomin and Sharma, Puru and Kang, Qingxuan and Jevdjic, Djordje and Deng, Junbo and Yang, Xingkun and Yu, Zhou and Zuo, Pengfei},
  booktitle={2024 USENIX Annual Technical Conference (USENIX ATC 24)},
  pages={111--126},
  year={2024}
}

@article{qiao2024conserve,
  title={ConServe: Harvesting GPUs for Low-Latency and High-Throughput Large Language Model Serving},
  author={Qiao, Yifan and Anzai, Shu and Yu, Shan and Ma, Haoran and Wang, Yang and Kim, Miryung and Xu, Harry},
  journal={arXiv preprint arXiv:2410.01228},
  year={2024}
}

@inproceedings{frantar2025marlin_mixed_precision,
  title={Marlin: Mixed-precision auto-regressive parallel inference on large language models},
  author={Frantar, Elias and Castro, Roberto L and Chen, Jiale and Hoefler, Torsten and Alistarh, Dan},
  booktitle={Proceedings of the 30th ACM SIGPLAN Annual Symposium on Principles and Practice of Parallel Programming},
  pages={239--251},
  year={2025}
}

@inproceedings{reggiani2023mix_precision,
  title={Mix-GEMM: An efficient HW-SW architecture for mixed-precision quantized deep neural networks inference on edge devices},
  author={Reggiani, Enrico and Pappalardo, Alessandro and Doblas, Max and Moreto, Miquel and Olivieri, Mauro and Unsal, Osman Sabri and Cristal, Adri{\'a}n},
  booktitle={2023 IEEE International Symposium on High-Performance Computer Architecture (HPCA)},
  pages={1085--1098},
  year={2023},
  organization={IEEE}
}

@article{bai2023longbench,
  title={Longbench: A bilingual, multitask benchmark for long context understanding},
  author={Bai, Yushi and Lv, Xin and Zhang, Jiajie and Lyu, Hongchang and Tang, Jiankai and Huang, Zhidian and Du, Zhengxiao and Liu, Xiao and Zeng, Aohan and Hou, Lei and others},
  journal={arXiv preprint arXiv:2308.14508},
  year={2023}
}

@article{chen2024progressive_quant,
  title={Progressive Mixed-Precision Decoding for Efficient LLM Inference},
  author={Chen, Hao Mark and Tan, Fuwen and Kouris, Alexandros and Lee, Royson and Fan, Hongxiang and Venieris, Stylianos I},
  journal={arXiv preprint arXiv:2410.13461},
  year={2024}
}

@inproceedings{huang2021mixed_precision,
  title={Mixed precision quantization for ReRAM-based DNN inference accelerators},
  author={Huang, Sitao and Ankit, Aayush and Silveira, Plinio and Antunes, Rodrigo and Chalamalasetti, Sai Rahul and El Hajj, Izzat and Kim, Dong Eun and Aguiar, Glaucimar and Bruel, Pedro and Serebryakov, Sergey and others},
  booktitle={Proceedings of the 26th Asia and South Pacific Design Automation Conference},
  pages={372--377},
  year={2021}
}

@misc{whatmatters2025,
      title={What Matters in Transformers? Not All Attention is Needed}, 
      author={Shwai He and Guoheng Sun and Zheyu Shen and Ang Li},
      year={2024},
      eprint={2406.15786},
      archivePrefix={arXiv},
      primaryClass={cs.LG},
      url={https://arxiv.org/abs/2406.15786}, 
}

@inproceedings{dong2019hawq_mixed_precision,
  title={Hawq: Hessian aware quantization of neural networks with mixed-precision},
  author={Dong, Zhen and Yao, Zhewei and Gholami, Amir and Mahoney, Michael W and Keutzer, Kurt},
  booktitle={Proceedings of the IEEE/CVF international conference on computer vision},
  pages={293--302},
  year={2019}
}

@inproceedings{risso2022channel_mixed_precision,
  title={Channel-wise mixed-precision assignment for dnn inference on constrained edge nodes},
  author={Risso, Matteo and Burrello, Alessio and Benini, Luca and Macii, Enrico and Poncino, Massimo and Pagliari, Daniele Jahier},
  booktitle={2022 IEEE 13th International Green and Sustainable Computing Conference (IGSC)},
  pages={1--6},
  year={2022},
  organization={IEEE}
}

@article{vaswani2017attention,
  title={Attention is all you need},
  author={Vaswani, Ashish and Shazeer, Noam and Parmar, Niki and Uszkoreit, Jakob and Jones, Llion and Gomez, Aidan N and Kaiser, {\L}ukasz and Polosukhin, Illia},
  journal={Advances in neural information processing systems},
  volume={30},
  year={2017}
}

@article{jiang2024neo_swiftLLM,
  title={Neo: Saving gpu memory crisis with cpu offloading for online llm inference},
  author={Jiang, Xuanlin and Zhou, Yang and Cao, Shiyi and Stoica, Ion and Yu, Minlan},
  journal={arXiv preprint arXiv:2411.01142},
  year={2024}
}

@inproceedings {agrawal2023sarathi,
author = {Amey Agrawal and Nitin Kedia and Ashish Panwar and Jayashree Mohan and Nipun Kwatra and Bhargav Gulavani and Alexey Tumanov and Ramachandran Ramjee},
title = {Taming {Throughput-Latency} Tradeoff in {LLM} Inference with {Sarathi-Serve}},
booktitle = {18th USENIX Symposium on Operating Systems Design and Implementation (OSDI 24)},
year = {2024},
isbn = {978-1-939133-40-3},
address = {Santa Clara, CA},
pages = {117--134},
url = {https://www.usenix.org/conference/osdi24/presentation/agrawal},
publisher = {USENIX Association},
month = jul
}

@inproceedings{patel2024splitwise,
  title={Splitwise: Efficient generative llm inference using phase splitting},
  author={Patel, Pratyush and Choukse, Esha and Zhang, Chaojie and Shah, Aashaka and Goiri, {\'I}{\~n}igo and Maleki, Saeed and Bianchini, Ricardo},
  booktitle={2024 ACM/IEEE 51st Annual International Symposium on Computer Architecture (ISCA)},
  pages={118--132},
  year={2024},
  organization={IEEE}
}

@article{yao2023comprehensive_quant,
  title={A comprehensive study on post-training quantization for large language models},
  author={Yao, Zhewei and Li, Cheng and Wu, Xiaoxia and Youn, Stephen and He, Yuxiong},
  journal={arXiv preprint arXiv:2303.08302},
  year={2023}
}

@article{nagel2021white_paper_quantization,
  title={A white paper on neural network quantization},
  author={Nagel, Markus and Fournarakis, Marios and Amjad, Rana Ali and Bondarenko, Yelysei and Van Baalen, Mart and Blankevoort, Tijmen},
  journal={arXiv preprint arXiv:2106.08295},
  year={2021}
}

@article{ma2024affinequant,
  title={Affinequant: Affine transformation quantization for large language models},
  author={Ma, Yuexiao and Li, Huixia and Zheng, Xiawu and Ling, Feng and Xiao, Xuefeng and Wang, Rui and Wen, Shilei and Chao, Fei and Ji, Rongrong},
  journal={arXiv preprint arXiv:2403.12544},
  year={2024}
}

@article{wu2023fastserve,
  title={Fast distributed inference serving for large language models},
  author={Wu, Bingyang and Zhong, Yinmin and Zhang, Zili and Liu, Shengyu and Liu, Fangyue and Sun, Yuanhang and Huang, Gang and Liu, Xuanzhe and Jin, Xin},
  journal={arXiv preprint arXiv:2305.05920},
  year={2023}
}

@misc{dynamic-layer-selection,
  author       = {Theodore Glavas et al.},
  title        = {Dynamic layer selection in decoder-only transformers},
  url          = {https://neurips.cc/virtual/2024/106468},
  note         = {Accessed: Oct 13, 2025}
}

@inproceedings{
jiang2025tracing,
title={Tracing Representation Progression: Analyzing and Enhancing Layer-Wise Similarity},
author={Jiachen Jiang and Jinxin Zhou and Zhihui Zhu},
booktitle={The Thirteenth International Conference on Learning Representations},
year={2025},
url={https://openreview.net/forum?id=vVxeFSR4fU}
}

@article{transformer-layers-as-painters, title={Transformer Layers as Painters}, volume={39}, url={https://ojs.aaai.org/index.php/AAAI/article/view/34708}, DOI={10.1609/aaai.v39i24.34708}, abstractNote={Despite their nearly universal adoption for large language models, the internal workings of transformers are not well understood. We aim to better understand the impact of removing or reorganizing information throughout the layers of a pretrained transformer. Such an understanding could both yield better usage of existing models as well as to make architectural improvements to produce new variants. We present a series of empirical studies on frozen models that show that the lower and final layers of pretrained transformers differ from middle layers, but that middle layers have a surprising amount of uniformity. We further show that some classes of problems have robustness to skipping layers, running the layers in an order different from how they were trained, or running the layers in parallel. Our observations suggest that even frozen pretrained models may gracefully trade accuracy for latency by skipping layers or running layers in parallel.}, number={24}, journal={Proceedings of the AAAI Conference on Artificial Intelligence}, author={Sun, Qi and Pickett, Marc and Nain, Aakash Kumar and Jones, Llion}, year={2025}, month={Apr.}, pages={25219-25227} }

@misc{async_cuda_stream,
  author       = {{Steve Rennich}},
  title        = {CUDA C/C++ Streams and Concurrency},
  url          = {https://developer.download.nvidia.com/CUDA/training/StreamsAndConcurrencyWebinar.pdf},
  note         = {Accessed: May 15, 2025}
}

@misc{nvidia_gds,
  author       = {{NVIDIA}},
  title        = {{GPUDirect Storage: A Direct Path Between Storage and GPU Memory}},
  url          = {https://developer.nvidia.com/blog/gpudirect-storage/},
  note         = {Accessed: May 15, 2025}
}

@misc{tgi,
  author       = {{Hugging Face}},
  title        = {Text Generation Inference},
  url          = {https://huggingface.co/docs/text-generation-inference/en/index},
  note         = {Accessed: May 15, 2025}
}

@misc{azure_LLM_inference_trace,
  author       = {{Azure}},
  title        = {Azure LLM Inference Traces},
  url          = {https://github.com/Azure/AzurePublicDataset/blob/master/AzureLLMInferenceDataset2023.md},
  year         = {2024},
  note         = {Accessed: May 15, 2025}
}

@misc{vicuna_7b_v1.5,
  author       = {{lmsys}},
  title        = {vicuna-7b-v1.5},
  url          = {https://huggingface.co/lmsys/vicuna-7b-v1.5},
  note         = {Accessed: May 15, 2025}
}

@misc{swiftllm_github_repo,
  author       = {{shengyu Liu and Jover Qian}},
  title        = {SwiftLLM},
  url          = {https://github.com/interestingLSY/swiftLLM},
  note         = {Accessed: May 15, 2025}
}

@misc{torchserve,
  author       = {{PyTorch}},
  title        = {Serve, optimize and scale PyTorch models in production},
  url          = {https://github.com/pytorch/serve},
  year         = {2023},
  note         = {Accessed: May 15, 2025}
}

@misc{triton,
  author       = {{NVIDIA Corporation}},
  title        = {Triton inference server: An optimized cloud and edge inferencing solution.},
  url          = {https://github.com/triton-inference-server/server},
  year         = {2019},
  note         = {Accessed: May 15, 2025}
}

@misc{fast_transformer,
  author       = {{NVIDIA Corporation}},
  title        = {FasterTransformer: Transformer related optimization, including BERT, GPT.},
  url          = {https://github.com/NVIDIA/FasterTransformer},
  year         = {2019},
  note         = {Accessed: May 15, 2025}
}
\bibliographystyle{mlsys2025}

\newpage
\section*{Appendix Overview}
\label{sec:appendix}
\begin{itemize}
    \item Section A: Layer Swapping Sequence Profiling
    \item Section B: Additional Experimental Results
    \item Section C: Implementation Details
    \item Section D: Limitations and Broader Impacts
\end{itemize}

\section*{A \quad Layer Swapping Sequence Profiling}
\label{sec:layer_profiling}

{\system} supports an optional offline profiling for the layer swapping sequence to further preserve model accuracy during runtime layer adaptation. This process consists of two key components: 
The \emph{Layer Importance Score (LIS)}, which ranks layers by combining the individual characteristics of each layer and its cumulative impact on overall model output;
and a \emph{greedy selection policy} that constructs the layer swapping sequence used during inference.

\subsection*{A.1 \quad Layer Importance Score (LIS)}
\label{appendix:LIS}

\noindent\textbf{Motivation and Design.}
The \emph{Layer Importance Score (LIS)} is defined as:
\begin{equation}
\text{LIS}_p = \alpha_1 \cdot \text{LTS}_p + \alpha_2 \cdot \text{LRS}_p + \beta \cdot \text{MDS}_p^{(Q)}
\label{eq:LIS}
\end{equation}

Here, $p$ indexes the candidate layer, $Q$ denotes the current set of quantized layers, and $\alpha_1$, $\alpha_2$, and $\beta$ are weighting coefficients.
LIS 
combines both \emph{layer-level} and \emph{model-level} sensitivity metrics to achieve accurate and generalizable layer ranking. Layer-level sensitivity captures how critical a single layer is by evaluating the degree of change between its input and output,
and quantization distortion. However, relying solely on layer-level metrics may lead to locally optimal sequences that ignore the cumulative impact on model output.
In contrast, model-level sensitivity measures the overall accuracy degradation introduced by swapping a given layer within the current model state. While this provides global awareness, depending exclusively on it risks overfitting to the specific calibration dataset.
To balance generality and robustness, LIS combines both global and local sensitivity metrics, \emph{without relying on backpropagation or reconstruction}.
This design ensures that the resulting layer swapping sequence preserves model accuracy while remaining data-agnostic and transferable across workloads. 

\noindent\textbf{Hyperparameter Selection.}
To balance robustness and accuracy, we define the hyperparameters of LIS as follows:
{$\alpha_1$} captures weight sensitivity (e.g., norm-based changes after quantization), {$\alpha_2$} measures quantization distortion in activation space (e.g., cosine similarity), and {$\beta$} reflects model-level degradation (e.g., perplexity under greedy layer removal).

\begin{table*}[h]
\caption{Perplexity comparison of LIS under different ($\alpha_1$,$\alpha_2$,$\beta$) hyperparameter settings across quantization levels on Llama 2 7B using WikiText-2.}
\label{table:perp_comp_lis}
\centering
\begin{tabular}{|c|c|c|c|c|c|c|}
\hline
\# Layers Quantized & 1 & 2 & 4 & 8 & 16 & 32 (Fully Quantized) \\ \hline
LIS (0.33,0.33,0.33) & \textbf{5.4732} & 5.4748 & 5.4786 & 5.4905 & 5.5257 & \textbf{5.6002} \\ \hline
LIS (0.25,0.25,0.5) & \textbf{5.4732} & \textbf{5.4743} & \textbf{5.4779} & \textbf{5.4875} & \textbf{5.5215} & \textbf{5.6002} \\ \hline
\end{tabular}
\end{table*}

We empirically choose $\alpha_1=\alpha_2=0.25$ and $\beta=0.5$, ensuring $\alpha_1 + \alpha_2 + \beta = 1.0$. This heuristic gives slightly more emphasis to {$\beta$} due to its stability across inputs and its direct alignment with user-perceived output quality. Relying only on {$\beta$}, however, risks overlooking layers with high local distortion, while {$\alpha_1$} and {$\alpha_2$} alone are more sensitive to input variance.
We compare this configuration to a baseline with uniform weighting ($\alpha_1=\alpha_2= \beta = 0.33$). As shown in the \hyperref[table:perp_comp_lis]{Table~\ref{table:perp_comp_lis}}, our setting produces lower perplexity across quantization levels and leads to more stable LIS rankings.

\noindent\textbf{Cosine Similarity.}  
LIS adopts cosine similarity as a lightweight and stable proxy for semantic drift during profiling. All sensitivity metrics are derived by quantifying the directional change between intermediate or final representations before and after layer morphing. Specifically, the cosine similarity between two vectors $\mathbf{a}$ and $\mathbf{b}$ is computed as: \vspace{-5pt}
\begin{equation}
\cos(\mathbf{a}, \mathbf{b}) = \frac{\mathbf{a} \cdot \mathbf{b}}{\|\mathbf{a}\| \, \|\mathbf{b}\|}
\label{eq:cosine}
\end{equation}
A higher similarity indicates smaller representational deviation and thus lower sensitivity to swapping.

\begin{table*}[h]
\caption{Perplexity comparison of LIS using L2 Norm vs. cosine similarity across quantization levels on
Llama 2 7B with WikiText-2.}\label{table:cosine_l2}
\centering
\begin{tabular}{|c|c|c|c|c|c|c|}
\hline
\# Layers Quantized & 1 & 2 & 4 & 8 & 16 & 32 (Fully Quantized) \\ \hline
LIS - L2 Norm & 5.4733 & 5.4776 & 5.4848 & 5.5021 & 5.5321 & \textbf{5.6002} \\ \hline
LIS - Cosine Similarity & \textbf{5.4732} & \textbf{5.4743} & \textbf{5.4779} & \textbf{5.4875} & \textbf{5.5215} & \textbf{5.6002} \\ \hline
\end{tabular}
\end{table*}

\noindent\textbf{Interpreting Layer-level and Model-level Sensitivity.} 
\textit{Layer Transformation Sensitivity (LTS)} measures the angular distance between a layer's input and output. A high LTS indicates weak transformation, suggesting the layer contributes minimally to representation learning.  
\textit{Layer Replacement Sensitivity (LRS)} quantifies the similarity between the outputs of the full-precision and quantized versions of the same layer. A high LRS implies low distortion and minimal risk of quality degradation upon replacement. Both LTS and LRS are computed independently of model outputs and remain consistent across input samples, making them robust to dataset shifts.  
\textit{Model Degradation Sensitivity (MDS)} captures the similarity between model outputs with and without a candidate layer replaced, conditioned on the current swapped set $Q$. A high MDS indicates minimal incremental impact when replacing the layer in context. MDS preserves overall model accuracy during sequential morphing.  
By combining local (LTS, LRS) and global (MDS) sensitivity metrics, {\system} avoids overfitting to specific calibration and achieves generalizable layer importance rankings across diverse datasets.

\subsection*{A.2 \quad Greedy Selection Policy}
Due to the combinatorial complexity of searching for the optimal layer swapping order, {\system} adopts a heuristic greedy strategy that incrementally constructs the morphing sequence based on the Layer Importance Score (LIS).

\noindent\textbf{Profiling Setting.}
Following the setup in~\cite{lin2024duquant, shao2023omniquant, ma2024affinequant}, we use a small calibration subset from the WikiText2 dataset, with the sequence length of 2,048. While LIS incorporates static layer-level metrics and model-level output feedback, its design avoids overfitting the specific calibration set. Once computed, the LIS ranking is fixed and reused across deployments, requiring no online re-tuning.

\noindent\textbf{Greedy Selection.}
{\system} employs a greedy selection policy guided by sensitivity metrics to construct an effective layer morphing sequence. The goal is to minimize cumulative degradation by progressively swapping the least impactful layers based on the LIS. 

\begin{algorithm}[t]
\caption{Swapping Sequence Profiling Based on Layer Importance Scoring (LIS)}
\label{alg:lis}
\begin{algorithmic}[1]
\Require Full-precision model $\mathcal{M}$, quantized model $\mathcal{M}^Q$, calibration dataset $\mathcal{D}$, params $(\alpha_1, \alpha_2, \beta)$
\For{each layer $i$}
    \State Compute $\text{LTS}_i = \text{CosSim}(\text{Input}_i, \text{Output}_i)$
    \State Compute $\text{LRS}_i = \text{CosSim}(\text{Output}_i, \text{Output}_i^Q)$
\EndFor
\State Initialize set of quantized layers $Q \gets \emptyset$
\For{$t = 1$ to $L$}
    \For{each unquantized layer $j \notin Q$}
        \State Temporarily quantize layer $j$ and evaluate model outputs
        \State Compute $\text{MDS}_j^{(Q)} = \text{CosSim}(f^{(Q)}(x), f^{(Q \cup \{j\})}(x))$
        \State Compute $\text{LIS}_j = \alpha_1 \cdot \text{LTS}_j + \alpha_2 \cdot \text{LRS}_j + \beta \cdot \text{MDS}_j^{(Q)}$
    \EndFor
    \State Select $j^* = \arg\max_j \text{LIS}_j$
    \State Add $j^*$ to $Q$ and replace corresponding layer in $\mathcal{M}$
\EndFor
\Return Ordered layer swap sequence $Q$
\end{algorithmic}
\end{algorithm}

As shown in \hyperref[alg:lis]{Algorithm~\ref{alg:lis}}, we first compute two input-independent metrics, LTS and LRS, for each layer using a small calibration dataset. Then, in each iteration, the algorithm evaluates every candidate layer by computing its MDS, conditioned on the current quantized set $Q$. The layer with the highest LIS is selected, added to $Q$, and swapped into the model. This process continues until all layers are ranked. 
The final sequence is fixed and reused at runtime.

\subsection*{A.3 \quad Evaluation}\label{appendix:eval}

To evaluate the effectiveness and generalizability of the LIS-based layer selection strategy, we compare it against several ordering baselines using perplexity across four models: Vicuna~\cite{vicuna_7b_v1.5}, Llama 2~\cite{touvron2023llama2}, Llama 3~\cite{grattafiori2024llama3}, and CodeLlama~\cite{roziere2023codellama}. 
The comparison includes the following baselines:
\emph{Front-to-Back}---layers are swapped sequentially from the input (first layer) to the output (last layer);
\emph{Back-to-Front}---the reverse order, starting from the final layer and proceeding backward; 
\emph{Random}---a randomly shuffled layer order, averaged over multiple runs to reduce variance. 
\vspace{-5pt}

\begin{table*}[ht]
\centering
\small
\caption{Perplexity results on WikiText2 under different layer swapping strategies for Vicuna 7B v1.5, Llama 2 7B, Llama 3 8B, and CodeLlama 34B. Each method is evaluated as the number of quantized (INT4) layers increases from 0 (fully FP16) to 32 or 48 (fully INT4).} \label{tab:perp_results_on_wikitext2}
\label{table:ppl}
\begin{tabular}{cc|ccccccc}
\toprule
\multirow{2}{*}{\textbf{Model}} & \multirow{2}{*}{\textbf{Method}} & \multicolumn{7}{|c}{\textbf{\# Swapped Decoder Layer}} \\
 & & \multicolumn{1}{|c}{\textbf{0 (FP16)}} & \multicolumn{1}{c}{\textbf{1}} & \multicolumn{1}{c}{\textbf{2}} & \multicolumn{1}{c}{\textbf{4}} & \multicolumn{1}{c}{\textbf{8}} & \multicolumn{1}{c}{\textbf{16}} & \multicolumn{1}{c}{\textbf{32 (INT4)}} \\
\midrule
\multirow{4}{*}{Vicuna 7B} 
& Front-to-Back & \multirow{4}{*}{\textbf{6.78}} & 6.79 & \textbf{6.78} & \textbf{6.79} & 6.80 & \textbf{6.84} & \multirow{4}{*}{\textbf{6.98}} \\
& Back-to-Front &                       & 6.82 & 6.83 & 6.84 & 6.86 & 6.91 &            \\
& Random        &                       & \textbf{6.78} & 6.79 & 6.80 & 6.82 & 6.87 &            \\
& LIS (ours)           &                       & \textbf{6.78} & \textbf{6.78} & \textbf{6.79} & \textbf{6.79} & \textbf{6.84} &            \\
\midrule
\multirow{4}{*}{Llama 2 7B} 
& Front-to-Back & \multirow{4}{*}{\textbf{5.47}} & \textbf{5.47} & \textbf{5.47} & \textbf{5.48} & 5.50 & 5.54 & \multirow{4}{*}{\textbf{5.60}} \\
& Back-to-Front &                       & 5.48 & 5.48 & 5.49 & 5.50 & 5.53 &            \\
& Random        &                       & \textbf{5.47} & 5.48 & \textbf{5.48} & 5.50 & 5.53 &            \\
& LIS (ours)           &                       & \textbf{5.47} & \textbf{5.47} & \textbf{5.48} & \textbf{5.49} & \textbf{5.52} &            \\
\midrule
\multirow{4}{*}{Llama 3 8B} 
& Front-to-Back & \multirow{4}{*}{\textbf{6.14}} & \textbf{6.15} & 6.16 & 6.19 & 6.23 & 6.34 & \multirow{4}{*}{\textbf{6.53}} \\
& Back-to-Front &                       & 6.17 & 6.18 & 6.20 & 6.24 & 6.33 &            \\
& Random        &                       & \textbf{6.15} & 6.16 & \textbf{6.18} & 6.24 & 6.34 &            \\
& LIS (ours)          &                       & \textbf{6.15} & \textbf{6.15} & \textbf{6.18} & \textbf{6.22} & \textbf{6.32} &            \\
\midrule
\multirow{4}{*}{CodeLlama 34B}
& Front-to-Back & \multirow{4}{*}{\textbf{5.47}} & \textbf{5.47} & \textbf{5.47} & 5.48 & \textbf{5.48} & \textbf{5.49} & \multirow{4}{*}{\textbf{5.53}} \\
& Back-to-Front &                       & \textbf{5.47} & 5.48 & 5.48 & \textbf{5.48} & \textbf{5.49} &            \\
& Random        &                       & \textbf{5.47} & \textbf{5.47} & 5.48 & \textbf{5.48} & \textbf{5.49} &       \multirow{2}{*}{\textbf{(48 INT4)}}     \\
& LIS (ours)          &                       & \textbf{5.47} & \textbf{5.47} & \textbf{5.47} & \textbf{5.48} & \textbf{5.49} &            \\
\bottomrule
\end{tabular}
\end{table*}

As shown in \hyperref[table:ppl]{Table~\ref{table:ppl}}, the LIS-based greedy selection strategy achieves strong and consistent perplexity results across all models and layer swapping levels, outperforming or matching heuristic baselines. Notably, the Front-to-Back strategy remains highly competitive, likely due to the model's ability to correct errors introduced in early swapped layers, making them safer to morph first. Due to its simplicity and effectiveness, {\system} adopts Front-to-Back as the default swapping policy when deploying new models or offline profiling is unavailable.

The LIS-based profiling is an optional, offline process that requires no runtime computation. For a 32-layer model, generating the full LIS sequence takes under 15 minutes on a single GPU. The process is efficiently parallelizable and only needs to be performed once per model. Once computed, the LIS ranking is reused during inference without incurring any runtime performance overhead. This design ensures that profiling enhances accuracy without sacrificing {\system}’s practicality in large-scale, latency-sensitive deployments.

\section*{B \quad Additional Experimental Results}

\subsection*{B.1 \quad Independence of Layer Quantization Effects}
\label{appendix:layer_independence}

To demonstrate the independence of layer quantization effects, we conducted a direct validation using Llama 2 7B on the WikiText dataset. We progressively quantized the first $N$ layers ($N = 0, 1, 2, 4, 8, 16$) and measured the additional perplexity increase by further quantizing either layer 19 or layer 24. As shown in \hyperref[tab:independence_of_layer]{Table~\ref{tab:independence_of_layer}}, the added perplexity from quantizing layer 19 remains consistently around 0.0035, regardless of how many earlier layers were already quantized. A similar pattern holds for layer 24 (0.0018-0.0020). This result provides support for the independence of layer-wise quantization effects.

\begin{table*}[h]
\caption{Evaluating independence of layer-wise quantization effects on perplexity under different quantization conditions.}
\label{tab:independence_of_layer}
\centering
\resizebox{\textwidth}{!}{%
\begin{tabular}{|c|c|c|c|c|c|c|}
\hline
\textbf{Quantized Layers} & \textbf{None} & \textbf{First 1 Layer} & \textbf{First 2 Layers} & \textbf{First 4} & \textbf{First 8} & \textbf{First 16} \\ \hline
PPL & 5.472089& 5.473293& 5.474974& 5.484205& 5.503622 &5.539723 \\ \hline
PPL (+ layer 19 quant.) & 5.475612 &5.476802& 5.478512& 5.487739& 5.507146& 5.543575 \\ \hline
\textbf{Effect (layer 19)} & \textbf{0.003523} & \textbf{0.003509} & \textbf{0.003538} & \textbf{0.003533} & \textbf{0.003524} & \textbf{0.003852} \\ \hline
PPL (+ layer 24 quant.) & 5.473969 & 5.475141 & 5.476794 & 5.486121 & 5.505570 & 5.541744 \\ \hline
\textbf{Effect (layer 24)} & \textbf{0.001880} & \textbf{0.001848} & \textbf{0.001820} & \textbf{0.001916} & \textbf{0.001948} & \textbf{0.002021} \\ \hline
\end{tabular}
}
\end{table*}

\subsection*{B.2 \quad A Verbatim Example for Mixed-Precision Serving}\label{appendix:verbatim_example}

{We further conducted a verbatim experiment to show {\system}'s mixed-precision performance using Llama 3 8B Instruct, generating 128 output tokens under different serving scenarios with the following input prompt: \textit{``Zebras are primarily grazers and can subsist on lower-quality vegetation. They are preyed on mainly by lions, and typically flee when threatened but also bite and kick.''}}

{
\begin{itemize}
    \item Response (FP16): That's correct! Zebras are indeed primarily grazers, and they are able to survive on a diet of lower-quality vegetation, such as grasses and shrubs. This is because they have a specialized digestive system...
    \item Response (Mixed-serving case 1): That's correct! Zebras are indeed primarily grazers (FP16), and they are well adapted to survive on a diet of grasses, leaves (W4), and other low-quality vegetation. Their digestive system is designed (FP16)...
    \item Response (Mixed-serving case 2): That's correct! Zebras are indeed primarily grazers (W4), and they are able to survive on a diet of lower-quality vegetation (FP16). Their digestive system is specially designed to break down and extract nutrients (W4)...
    \item Response (W4): That's correct! Zebras are indeed primarily grazers, and they are well adapted to survive on a diet of grasses, leaves, and other low-quality vegetation. Their digestive system is designed to break down...
\end{itemize}
}

{{\system}'s mixed-precision serving allocates compute budget adaptively, ensuring robust performance without compromising user experience.}

\subsection*{B.3 \quad Effectiveness Across Quantization Algorithms}\label{appendix:effective_across_quant}

{\hyperref[tab:perf_ablation_vicuna]{Table~\ref{tab:perf_ablation_vicuna}} shows the performance of {\system} with AWQ and Uniform INT4 quantization vs. FP16 on Vicuna 7B v1.5, using the QMSum dataset and Azure trace.}

\begin{table*}[h]
\caption{Performance of {\system} with AWQ and Uniform INT4 quantization vs. FP16 on Vicuna 7B v1.5, using the QMSum dataset and Azure trace.} \label{tab:perf_ablation_vicuna}
\centering
\begin{tabular}{|c|c|c|c|c|}
\hline
Quantization (INT4) & Metric & Static Quantization & {\system} & Full Precision \\
\hline
& TTFT P95 (s) & 0.4086 & 0.7281 & 6.2376 \\
\cline{2-5}
AWQ & TPOT P99 (s) & 0.0846 & 0.1123 & 0.1384 \\
\cline{2-5}
& F1 Score & 11.77 & 12.85 & 13.01 \\
\hline
& TTFT P95 (s) & 0.3958 & 0.7034 & 6.2376 \\
\cline{2-5}
Uniform & TPOT P99 (s) & 0.0820 & 0.1095 & 0.1384 \\
\cline{2-5}
& F1 Score & 11.32 & 12.57 & 13.01 \\
\hline
\end{tabular}
\end{table*}

\section*{C \quad Implementation Details}
\label{sec:implementation}
\noindent\textbf{Implementation.}
{\system} is built on top of SwiftLLM~\cite{swiftllm_github_repo, jiang2024neo_swiftLLM}, with approximately \emph{2,200 lines of Python} and \emph{500 lines of C++/CUDA}. It adds runtime support for dynamic layer swapping and elastic KVC resizing with minimal changes to the scheduler and attention mechanisms, such as {\FlashAttention}~\cite{dao2022flashattention, dao2023flashattention_v2} and {\PagedAttention}~\cite{kwon2023efficient_vllm}, and remains compatible with state-of-the-art LLM inference engines such as vLLM~\cite{kwon2023efficient_vllm}.
At initialization, full-precision and quantized transformer layer weights (\emph{FP16, W8, W4}) are preloaded into pinned CPU memory, and all \emph{GEMM} kernels are precompiled using dummy data to eliminate runtime compilation overhead. GPU memory regions for each layer are preallocated, enabling in-place weight swapping via \emph{cudaMemcpyAsync} without pointer remapping. For KV cache resizing, we extend {\PagedAttention} to support block-level reallocation and remapping through dynamic memory registration. Morphing and decoding are executed on separate CUDA streams to ensure efficient asynchronization and minimize interference with token generation. 

\noindent\textbf{Experiment Settings.} 
We evaluate {\system} on four representative open-weight LLMs: Vicuna 7B, Llama 2 7B, Llama 3 8B, and CodeLlama 34B. Models with 7B/8B parameters are run on NVIDIA L4 GPUs (24~GB),
while the 34B model is evaluated on an NVIDIA A100 GPU (80~GB). For models using Multi-Head Attention (MHA), we set context lengths to 512 for prompts and 256 for responses; for Grouped-Query Attention (GQA) models, we use 1024/512. All models are loaded with pre-quantized AWQ INT4 weights.

\noindent\textbf{Serving Traces.}
The \textbf{Azure LLM Inference Dataset 2023 trace}~\cite{azure_LLM_inference_trace, patel2024splitwise} is a publicly released dataset capturing anonymized LLM request logs from Azure’s cloud infrastructure. It includes request arrival times, prompt, and output lengths statistics. The dataset is publicly available at \href{https://github.com/Azure/AzurePublicDataset}{\url{https://github.com/Azure/AzurePublicDataset}}. For our evaluation, we sample 72 seconds of traffic with a downscaling factor of $4.75\times$ to match the hardware memory footprint and enable simulation of large-batch request bursts. 
\textbf{BurstGPT}~\cite{wang2024burstgpt} is a real-world LLM inference workload trace collected from a university campus. It captures naturally occurring burst patterns resulting from student and faculty interactions with deployed chatbots and LLM-based tools. The trace includes detailed request metadata such as arrival timestamps, prompt lengths, and session-level characteristics, enabling realistic simulation of latency-sensitive serving conditions. It is publicly available at \href{https://github.com/HPMLL/BurstGPT}{\url{https://github.com/HPMLL/BurstGPT}}. For our evaluation, we also extract a 72-second segment and apply a $1.75\times$ downscaling factor to simulate saturation-level conditions. This trace is used to benchmark {\system}’s responsiveness and adaptation under real-world burst traffic.

\noindent\textbf{Evaluation Datasets.}
\textbf{GovReport}~\cite{gov_report} is a long-form summarization dataset consisting of U.S. government reports paired with expert-written summaries. It is publicly available at \href{https://huggingface.co/datasets/launch/gov_report}{\url{https://huggingface.co/datasets/launch/gov\_report}}. We use it to benchmark summarization quality and stress-test long input handling. Average document length exceeds 2,000 tokens, making it suitable for evaluating memory-intensive generation.
\textbf{QMSum}~\cite{zhong2021qmsum} is a query-based meeting summarization dataset comprising multi-party meeting transcripts with user-specified queries and corresponding abstractive summaries. Available at \href{https://github.com/Yale-LILY/QMSum}{\url{https://github.com/Yale-LILY/QMSum}}, it tests both summarization and task-oriented comprehension under long-context inputs.
\textbf{DuReader}~\cite{he2017dureader} is a Chinese machine reading comprehension dataset with over human-annotated question-answer pairs from Baidu search logs. It covers open-domain QA with a range of answer formats. We use the English-translated version and evaluate factual correctness. The dataset is hosted at \href{https://github.com/baidu/DuReader}{\url{https://github.com/baidu/DuReader}}.
\textbf{Multi-News}~\cite{fabbri2019multi} is a multi-document summarization dataset containing news articles from multiple sources clustered around the same event, with human-written summaries. It is accessible at \href{https://github.com/Alex-Fabbri/Multi-News}{\url{https://github.com/Alex-Fabbri/Multi-News}}. This dataset evaluates the model’s ability to synthesize content across multiple documents and is especially useful for benchmarking performance on broad-context summarization.

To construct realistic evaluation workloads, we align request arrival traces--which provide only timestamps and arrival rates--with benchmark datasets that contain task-specific input contexts but no temporal information, pairing each incoming request with a sampled context to form a complete sequence of timestamped, content-rich requests.

\section*{D \quad Limitations and Broader Impacts}
\label{sec:limit_impact}

\noindent\textbf{Limitations.}
While {\system} is practical and effective for dynamic LLM serving, several limitations remain.
To support runtime layer swapping, {\system} stores both full-precision and quantized variants in host memory. Although this increases memory usage, the overhead is typically under $2\times$ the model size and is well accommodated by modern LLM serving clusters. Future work may further reduce this cost by streaming layers from SSD to host memory on demand or directly fetching them from SSD via GPUDirect Storage (GDS)~\cite{nvidia_gds}.
{\system} currently applies morphing at the transformer layer level. While effective, finer-grained adaptation, such as independently adjusting attention and MLP submodules, could unlock additional efficiency and precision flexibility.
{\system} reacts to system pressure in real time but does not anticipate upcoming surges. Integrating lightweight workload forecasting could enable proactive morphing decisions and further improve responsiveness under bursty traffic.

\noindent\textbf{Broader Impacts.}
{\system} is designed to improve the efficiency and elasticity of LLM serving under real-world, dynamic workloads. Its ability to reduce tail latency and alleviate memory pressure during high-traffic scenarios enhances the responsiveness and accessibility of language models, especially in environments with constrained compute resources such as edge devices or public-serving infrastructures. By allowing runtime trade-off navigation between accuracy and latency, {\system} enables system designers to align inference behavior with user-facing service priorities, such as delivering faster responses for interactive applications, without requiring permanently quantized models or over-provisioned compute clusters. This flexibility supports broader deployment of LLMs across diverse platforms and use cases, contributing to the democratization of AI capabilities.

\end{document}